\documentclass[11pt]{article}

\usepackage{amssymb,amsmath,latexsym,bm,amsfonts}
\usepackage{graphicx}
 
\usepackage{longtable}
\usepackage{color,xcolor}
\usepackage{indentfirst}
\usepackage{subfigure}
\usepackage[margin=1in]{geometry}
\usepackage{mathtools,tensor,graphicx,amsmath,amssymb}
\numberwithin{equation}{section}

\usepackage[nottoc ]{tocbibind}

\newcommand{\be}{\begin{equation}}
\newcommand{\ee}{\end{equation}}
\newcommand{\bpm}{\begin{pmatrix}}
\newcommand{\epm}{\end{pmatrix}}

\newcommand{\PBK}[1]{\ensuremath{\begin{pmatrix}#1\end{pmatrix}}}

\newcommand{\EV}[1]{\langle #1 \rangle}
\newcommand{\beqn}{\begin{eqnarray}}
\newcommand{\eeqn}{\end{eqnarray}}

\newcommand{\cD}{\mathcal D}

\newcommand{\ta}{\theta}
\newcommand{\taa}{\theta\theta}

\newcommand{\tab}{\bar\theta }
\newcommand{\taab}{\bar\theta\bar\theta}

\newcommand{\p}{\partial}

\DeclareMathOperator{\sgn}{sgn}
\DeclareMathOperator{\Real}{Re}

\DeclareMathOperator{\Imag}{Im}

\newcommand{\sfD}{\textsf D}
\newcommand{\sfF}{\textsf F}
\usepackage{breqn}
\usepackage[colorlinks=true,filecolor=red,citecolor=blue ]{hyperref}

\usepackage{eufrak}
 \usepackage{mathrsfs}
 \usepackage{mathrsfs}
\begin{document}

\renewcommand{\baselinestretch}{1.2}

\begin{titlepage}
  \ \\
  \vskip 1cm

  \begin{center}

    \Large{  $\mathcal N=2$ Supersymmetry Deformations, Electromagnetic Duality and Dirac-Born-Infeld Actions}

    \vspace{0.8 cm}
    \normalsize{Ignatios Antoniadis$^{a,b}$, Hongliang  Jiang$^{b}$  and Osmin Lacombe$^{a}$ 
\let\thefootnote\relax\footnote{Email: antoniadis@itp.unibe.ch, jiang@itp.unibe.ch, osmin@lpthe.jussieu.fr}
\addtocounter{footnote}{-1}\let\thefootnote\svthefootnote  
    }
    
    \vspace{1cm}
    \normalsize{\it  $^{a}$Laboratoire de Physique Th\'eorique et Hautes Energies - LPTHE \\
Sorbonne Universit\'e, CNRS, 4 Place Jussieu, 75005 Paris, France \\ [10pt]
$^{b}$Albert Einstein Center, Institute for Theoretical Physics, University of Bern, \\
 			Sidlerstrasse 5, 3012 Bern, Switzerland } \\

    \vspace{0.8 cm}

    \begin{abstract}  
We study the general deformation of $\mathcal N=2$ supersymmetry transformations of a vector multiplet that forms a (constant) triplet under the $SU(2)$ R-symmetry corresponding to the magnetic dual of the triplet of the Fayet-Iliopoulos (FI) parameters. We show that in the presence of both triplets, the induced scalar potential of a vector multiplet with generic prepotential has always a minimum that realises partial breaking of $\mathcal  N=2\to \mathcal  N=1$ supersymmetry. We then consider the impact of the deformation in the Dirac-Born-Infeld (DBI) action where one supersymmetry  is non-linearly realised, described by a nilpotent constraint on the deformed $\mathcal N=2$ chiral-chiral superfield. We show that the generic magnetic deformation induces an ordinary FI D-term along the linear supersymmetry via the theta-angle. Moreover, we argue that the resulting action differs on-shell from the standard one (DBI+FI) by fermionic contributions.

    \end{abstract}

    \vspace{1cm}
      
  \end{center}
  
\end{titlepage}

\pagestyle{plain}
\pagenumbering{arabic}

\tableofcontents

\newpage

\newpage

\section{Introduction}

Partial breaking of $\mathcal N=2$ global supersymmetry (SUSY) to $\mathcal N=1$ requires a deformation of supersymmetry transformations~\cite{APT, ADtM}. The latter consists in  adding arbitrary complex constants which modifies the transformations of fermions but leaves intact the supersymmetry algebra of infinitesimal transformations. Some of these constants can be absorbed by shifting the auxiliary fields and, thus, do not correspond to genuine deformations. One therefore expects that a general deformation contains the same number of parameters as the number of real auxiliary fields in every supersymmetry multiplet, consisting technically  {in} adding constant imaginary parts. Indeed, this is the case for $\mathcal N=2$ vector and single-tensor multiplets that can be deformed by adding three or two constant parameters, correspondingly~\cite{ADcM}. Partial supersymmetry breaking implies a special relation among the deformation parameters guaranteeing the existence of a linear combination of the two supersymmetries under which all fermions of the multiplet transform linearly (without constants).

 In this work, we study the general deformation of supersymmetry for $\mathcal N=2$ vector multiplets and its effect on two-derivative effective actions involving a generic prepotential, as well as  {on} the DBI action{,} where one supersymmetry is non-linearly realised. The general deformation forms a triplet under the $SU(2)_R$ symmetry and consists technically in adding a constant imaginary part to the triplet  of auxiliary  fields, formed by the (complex) F- and (real) D-auxiliary components of the $\mathcal N=1$ chiral and vector multiplet that compose the $\mathcal N=2$ double chiral vector ${\cal W}=(X,W)$. The deformation associated to $F$ is known to give rise to a magnetic Fayet-Iliopoulos (FI) term proportional to the special coordinate $f_X\equiv{\partial_X}f$ where $f(X)$ is the holomorphic $\mathcal N=2$ prepotential~\cite{APT}. Indeed, ordinary electromagnetic duality exchanges $X$ with $f_X$ and, thus, their corresponding coefficients. Here, we extend this result to the D-auxiliary whose deformation modifies the Bianchi identity of $W$ and we show that this modification is dual to the ordinary FI parameter under electromagnetic duality.
 
We then study the general two-derivative effective action of a deformed $\mathcal N=2$ double chiral  {multiplet} and show that it exhibits a partial $\mathcal N=2\to \mathcal N=1$ breaking at the minimum of the scalar potential for generic values of the parameter space. Special values may leave $\mathcal N=2$ unbroken or a runaway potential but  {one can never realize} complete breaking of both supersymmetries, unless trivially in a free theory. This result was expected since one could obtain it by   using  a  $SU(2)_R$ rotation from the cases studied  {in the literature}~\cite{APT, ADcM}. The analysis is however useful for unveiling the main properties of the D-deformation that will be relevant in the context of  Dirac-Born-Infeld (DBI) actions. Complete breaking {requires} at least two vector multiplets. For instance, in the simplest case, one can combine two independent theories, each one breaking $\mathcal N=2\to \mathcal N=1$ in a different direction. An interesting observation is that the D-deformation described above gives rise to an ordinary FI D-term proportional to the theta-angle. For a generic prepotential, this term is of course field dependent, while it becomes constant only in the free theory. 

We {next} extend our analysis to the case of the DBI action where one supersymmetry  is non-linearly realised, describing the effective field theory of a D3-brane in an $\mathcal N=2$ supersymmetric bulk~\cite{BG}. The deformation is now implemented in the nilpotent constrained deformed superfield~\cite{ADtM, ADcM} and  {we  find essentially} the same result as in the previous unconstrained case of a general prepotential. This time there is no scalar potential but the parameters of the FI term and the general deformation can be absorbed into a redefinition of the DBI couplings, namely the non-linear supersymmetry breaking scale (or the brane tension), the $U(1)$ gauge coupling and the theta-angle. We notice again that the D-deformation gives rise to an FI D-term through the theta-angle. This time the FI term is constant and the theory is not free. In principle, one would expect that the presence of this term would break both supersymmetries  but this  is not the case. Instead, one $\mathcal N=1$ linear supersymmetry remains but it changes direction. In the string theory context, it corresponds to rotate the brane in the bulk. As in the previous case, the complete breaking of supersymmetry can arise only in a system of at least two DBI actions preserving different linear {supersymmetries}, corresponding to two branes at angles.

Despite  the fact that the FI term induced by the D-deformation via the theta-angle gives the same bosonic action as adding a standard FI term to the DBI, the fermionic part of the action appears to be different~\cite{AJL} suggesting that this is yet another way to write a constant FI term in global supersymmetry, at least for $\mathcal N=2$ with one supersymmetry non-linearly realised.\footnote{Note that the new FI D-term proposed in~\cite{CFTvP} preserves only $\mathcal N=1$ supersymmetry.} The effective D-brane action was computed up to interaction terms of dimension-eight, or equivalently second order in the Regge slope $\alpha'$, and was compared with the expansion of the supersymmetric DBI action in~\cite{BBdRS}. It would be interesting to compute its modification in the presence of an FI D-term induced at the string level, for instance by internal magnetic fields, and compare with the different effective field theory actions. The coupling to supergravity is another interesting question, in particular whether it implies the gauging of the R-symmetry. Indeed, the absence of the extra fermionic contribution associated to the standard FI term exhibiting the gauging of the R-symmetry~\cite{CFTnonlinear} suggests that this gauging may not be necessary for the coupling to supergravity in our case.

The outline of our paper is the following. In Section~2, we review the general deformation of $\mathcal N=2$ supersymmetry transformations for a chiral-chiral multiplet and the condition for a partial $\mathcal N=2\to \mathcal N=1$ breaking.  In Section~3, we establish the electromagnetic  duality at fully $\mathcal N=2$ level. Adding deformations {is} shown to be equivalent to  {adding} the triplet of FI terms in the dual theory.  
In Section~4, we analyse the general $\mathcal N=2$ action based on an arbitrary deformed vector superfield; we compute the scalar potential and show that the only non-trivial minima break partially $\mathcal N=2\to \mathcal N=1$. In Section~5, we study the  {generalization} of the DBI action with the D-deformation and show that it leads to an  FI term via the theta-angle. We analyse its bosonic part and show that the deformation and FI parameters can be absorbed in the independent couplings, leaving the usual DBI form invariant. We also discuss the fermionic terms and argue that the FI term induced by the D-deformation is different from the standard FI term added to the DBI action. Section~6 contains some concluding remarks.

   
\section{General  deformations in $\mathcal N=2$ }  \label{sec:n2def} 

In this section, we investigate the properties of $\mathcal N=2$ vector multiplet.  {We then} consider the most general deformation of this vector multiplet which can be {parameterized} by three real constants. The deformation yields the non-linear realization of one supersymmetry.

\subsection{$\mathcal N=2$ vector multiplet: structure,   transformation and symmetry  }
 
We start  with the following chiral-chiral $\mathcal N=2$  multiplet  
\be\label{n2chiral}
\mathcal  W (y, \theta, \tilde \theta)= X(y,\theta )+\sqrt{2} i \tilde  \theta  W(y,\theta ) - \tilde \theta \tilde \theta G(y,\theta ), 
\qquad y^\mu=x^\mu+ i \ta \sigma ^\mu \tab+ i \tilde \ta \sigma ^\mu \bar{\tilde\theta}~,
\ee
which is chiral with respect to both supersymmetries
\footnote{We follow the conventions in \cite{Wess:1992cp}, so the superspace covariant  {derivatives}, in terms of the chiral coordinate, are given by 
\be
D_\alpha =\frac{\p}{\p \ta^\alpha}+2i \sigma^\mu_{\alpha \dot \alpha}\tab^{\dot \alpha} \frac{\p}{\p y^\mu},\qquad \bar D_{\dot\alpha} =-\frac{\p}{\p \bar\ta^{\dot\alpha}}
\ee
and similarly for $\tilde D_\alpha,  \bar{ \tilde{D}}_\alpha$.
}:
\be
\bar D\mathcal W= \bar{ \tilde{D}}\mathcal W=0~.
\ee
 {The fields transform as follows under the second supersymmetry}\footnote{The first supersymmetry refers to the supersymmetry associated with $\theta$, while the  second supersymmetry refers to the  one associated with $\tilde\theta$.}:
\beqn
\tilde \delta X&=&\sqrt2 i \epsilon W~, \\
\tilde \delta W &=&\sqrt2  \sigma^\mu \bar \epsilon  \p_\mu X+\sqrt2  i \epsilon G~, \\
\tilde \delta  G&=& - \sqrt2 \p_\mu W \sigma^\mu \bar \epsilon~.   \label{susytsfG}
\eeqn

The superfield \eqref{n2chiral} is reducible and describes the degrees of freedom of a $\mathcal N=2$ vector and tensor multiplet. To reduce them to those of a vector, one requires $W$ to be the field-strength superfield of a $\mathcal N=1$ vector multiplet, satisfying $DW-\bar D\bar W=0$.
Furthermore, one can verify explicitly that $\frac14 \bar D^2 \bar X$ transforms in the same way  {as  $G$} in \eqref{susytsfG}. 
Therefore 
we can set 
\be
G=\frac14 \bar D^2 \bar X
\ee
without violating the $\mathcal N=2$  supersymmetry.

Since $\mathcal W$ is chiral with respect to both supersymmetries, we can   consider the following   action
\be
\mathcal L_{\mathcal N=2\, \mathcal W^2+c.c} = \frac14\int d^2 \theta  d^2 \tilde \theta \mathcal W^2+c.c.=\frac14\int d^2 \theta  \Big( W^2 -2 XG \Big)+c.c. 
=\frac14 \int d^2 \theta (W^2-\frac12 X\bar D^2 \bar X)+c.c.~.\qquad
\ee
On the other hand, the $\mathcal N=2$  Maxwell theory, in terms of $\mathcal N=1$ language,  is described by a chiral multiplet $X$ and a vector multiplet $W$ with action    given by 
\be\label{N2Maxwell}
\mathcal L_{\mathcal N=2 \text{ Maxwell}}=\int d^2 \theta d^2 \bar \theta \bar X X +\frac14 \int d^2 \theta W^2+\frac14 \int d^2 \theta \bar W^2
=\frac14 \int d^2 \theta (W^2-\frac12 X\bar D^2 \bar X)+c.c.~,
\ee
up to a total derivative.
We see that the above two actions are  equivalent, implying that the extra constraint imposed on $\mathcal W$ is correct.  
 
 Thus  the  $\mathcal N=2$  vector multiplet can be described, in term of $\mathcal N=2$ superfield, as
\be\label{n2vector}
\mathcal  W (y, \theta, \tilde \theta)= X(y,\theta )+\sqrt{2} i \tilde  \theta  W(y,\theta ) -\frac14 \tilde \theta \tilde \theta  \bar D^2 \bar X(y,\theta)~,
\ee
where $X,W$ are  $\mathcal N=1$ chiral and vector  multiplets, respectively. {Their component forms read:}
\beqn
W_\alpha &=& -i \lambda_\alpha +  \theta_\alpha \sfD  -  i (\sigma^{\mu\nu}\theta)_\alpha F_{\mu\nu}    +\taa (\sigma^\mu   \p_\mu {\bar \lambda}  )_\alpha~,
\\
X&=&x+\sqrt{2} \theta \chi -\theta\theta \sfF~,
\\
\frac14 \bar D^2 \bar X&=&\bar \sfF- \sqrt{2}i \theta \sigma^\mu \p_\mu \bar \chi - \theta\theta \eta^{\mu  \nu}\p_\mu\p_\nu \bar x~.
\eeqn

Alternatively, the $\mathcal N=2$  vector multiplet \eqref{n2vector} can be obtained from \eqref{n2chiral} by imposing the following irreducibility conditions:
\be\label{irreducibleW}
\cD_i \cD_j \mathcal W =\epsilon_i{}^k\epsilon_j {}^l   \bar \cD_k \bar  \cD_l \bar{ \mathcal W }~, \qquad i,j,k,l=1,2~.
\ee
Here $\cD_1=D, \cD_2=\tilde D$ correspond to the super-covariant derivatives of the first and second supersymmetry.   The antisymmetric symbol is defined as $\epsilon_1{}^1=\epsilon_2{}^2=0, \epsilon_1{}^2=-\epsilon_2{}^1=1$.
From \eqref{n2vector}, we can read the transformation rules of $X$ and $W$ under the second supersymmetry
\beqn
\tilde \delta X&=&\sqrt 2  i \tilde \epsilon W~,  \\
\tilde \delta  W_\alpha&=&\sqrt 2 i \Big( \frac14 \tilde \epsilon_\alpha \bar D^2 \bar X - i (\sigma^\mu \bar{ \tilde \epsilon})_\alpha \p_\mu X  \Big) ~. 
\eeqn

We are especially interested in the  {auxiliary field part of the} SUSY transformation rules of fermions. 
Under the second supersymmetry, the fermions transform as
\beqn
\tilde \delta \lambda_\alpha  &=&-\sqrt2 \bar \sfF \tilde \epsilon_\alpha \nonumber~,  \\
\tilde \delta \chi_\alpha  &=& i \sfD\tilde  \epsilon_\alpha   ~,
\eeqn
while under the first supersymmetry, they transforms as
\beqn
 \delta \lambda_\alpha  &=& i\sfD  \epsilon_\alpha~,  \nonumber \\
  \delta \chi_\alpha  &=&-\sqrt2   \sfF \epsilon_\alpha  ~.
\eeqn
 The full SUSY transformation of the fermions can then be written as
 \be\label{fermiontransf}
 \delta_{susy} \PBK{\chi_\alpha \\ \lambda_\alpha}
 =\PBK{ -\sqrt2  \sfF & i \sfD\\ i\sfD  &-\sqrt 2 \bar \sfF  } 
 \PBK{\epsilon_\alpha \\ \tilde \epsilon_\alpha}~.
 \ee
 
The $\mathcal N=2$ vector multiplet  $\mathcal W$ has $SU(2)_R$ invariance. To see this symmetry, we define the following  $SU(2)_R$ doublets
\be
\vartheta^1=\theta~, \qquad \vartheta^2=\tilde \theta~, \qquad  \eta_1=\chi~, \qquad \eta_2=\lambda~.
\ee
 {The vector multiplet can be expanded in components as}
\be
 {\mathcal W} (y, \theta, \tilde \theta)=x+\sqrt2(\theta \chi+\tilde \theta \lambda )-\taa F-\tilde \theta\tilde \theta \tilde F+i\sqrt2 \theta\tilde \theta \sfD+...=
 x+\sqrt{2} \vartheta^i \eta_i -\vartheta^i\vartheta^j Y_{ij}+...~,
\ee
where 
\be
Y_{ij}=Y_{ji}=\Big( \bm Y \cdot \bm  \sigma \sigma_2 \Big)_{ij}~,
\ee
with   $\bm\sigma=(\sigma_1,\sigma_2,\sigma_3)$   the standard Pauli matrices. More explicitly,
\be\label{Yvev}
Y_{11}=\sfF~, \qquad Y_{22} =\bar \sfF~, \qquad Y_{12}=-  \frac{i}{\sqrt{2}} \sfD, \qquad \bm Y =\Big( \Imag \sfF, \Real \sfF, \frac{\sfD}{\sqrt{2}}\Big)~.
\ee

For convenience, we also construct the following triplet {of fermionic} coordinates transforming in the adjoint representation of $SU(2)_R$
\be
\bm\Theta=\PBK{\theta &\tilde \theta}\bm \sigma   \sigma_2  \PBK{\theta \\ \tilde \theta}
=\Big(  i( \taa - \tilde\ta\tilde \ta ), (\taa + \tilde\ta\tilde \ta ), - 2 i \ta \tilde\ta   \Big)~.
\ee
 {It can be used to form the quantity:} 
\be
\bm \Theta \cdot \bm Y =\taa \sfF+\tilde \theta\tilde \theta \bar \sfF - \sqrt2 i \theta \tilde \theta \sfD= \vartheta^i\vartheta^j Y_{ij}~.
\ee
Note that the   $SU(2)_R$ symmetry can also be seen from the  $SU(2)_R$ invariant reality conditions:
\be\label{N=2real}
 Y_{ij}^* =\epsilon_i{}^k  \epsilon_j{}^l Y_{kl}  ~.
\ee

\subsection{General deformation}

We are going to  {modify $\bm Y$ by adding a constant deformation $\bm Y_{\text{def}}$}. The real part  {of} $\bm Y_{\text{def}}$ can be absorbed to a trivial shift of the auxiliary fields in $\bm Y$. {Hence} we only need to focus on a pure imaginary $\bm Y_{\text{def}}$ \cite{ADcM}. Using the $SU(2)_R$ symmetry, we can rotate the vector $\bm Y_{\text{def}}$ to any specific direction. As we will see, this just {indicates}  that the model always has $\mathcal N=1$ residual supersymmetry  after deformation.  However, the direction of the residual supersymmetry depends on the deformation parameters  {which is} important for {the} purpose of total supersymmetry breaking. Therefore we  don't  rotate the deformation vector $\bm Y_{\text{def}}$ and consider the following generic deformation:
\be 
\bm Y_{\text{def}}=\Big(  i\frac{1}{4\kappa}  \cos \phi ,    i\frac{1}{4\kappa}  {\sin \phi},  i \frac{\gamma}{\sqrt2  }\Big)~,  \qquad \gamma, \phi, \kappa\in \mathbb R~.
\ee 
It contains three deformation parameters.  {As we said earlier the real part of the deformation vector has no physical effects, thus we can equivalently choose
 \be\label{Ydef}
\bm Y_{\text{def}}=\Big( \frac{i}{4\kappa} e^{i\phi},\frac{1}{4\kappa} e^{i\phi},  i\frac{\gamma}{\sqrt2 }\Big)~.  
\ee   
In the remainder of the paper, we will study the general  deformation in the form of  \eqref{Ydef}. 

 {The deformation $\bm Y_{\text{def}}$ induces a deformation ${\mathcal W}_{\text{def}}$ of the superfield $\mathcal W$. It reads}
\be\label{def1}
 {\mathcal W}_{\text{def}}= - \bm \Theta \cdot \bm Y_{\text{def}}= -\frac{1}{2\kappa} e^{i\phi}\tilde \theta\tilde \theta -\sqrt2\gamma  \theta \tilde \theta ~,
\ee 
 {and} modifies the irreducibility condition \eqref{irreducibleW}  to
 \footnote{Similar modification was obtained  in \cite{Ivanov:1997mt} through EM duality transformation which will be discussed in the next section.}
\be\label{defIrreducilbeW}
\cD_i \cD_j \mathcal W - \epsilon_i{}^k\epsilon_j {}^l   \bar \cD_k \bar  \cD_l \bar{ \mathcal W }= i \gamma_{ij}~, \qquad \gamma_{ij}\in \mathbb R~,
\ee
where
\be
\gamma_{ij}=8 \Big(  \Imag( \bm Y) \cdot \bm  \sigma \sigma_2 \Big)_{ij}~.
\ee 
In particular, this implies  {the following equation}\footnote{This modification appeared before in \cite{Kuzenko:2009ym}.}
\be\label{defW}
DW-\bar D\bar W=-4i \gamma~,
\ee  
which modifies the  standard Bianchi identity of $\mathcal N=1$ vector multiplets. The deformed vector multiplet can be solved and expressed in components  as
\be
W_\alpha = -i \lambda_\alpha +  \theta_\alpha \sfD  -  i (\sigma^{\mu\nu}\theta)_\alpha F_{\mu\nu}    +\taa (\sigma^\mu   \p_\mu {\bar \lambda}  )_\alpha~,
\ee
where 
\be
\sfD=d+i\gamma, \qquad d, \gamma\in \mathbb R~.
\ee
Here $\gamma$ is a constant and  $d$ is the auxiliary field that should be eliminated through its equation of motion. 
    
 \subsection{Deformed supersymmetric transformation and supersymmetry breaking}\label{susybreaking}
 
 {In order to} discuss supersymmetry transformations and supersymmetry  breaking,   {one} should take into account both the deformations and the dynamical parts sourced by the auxiliary fields.   {It is} convenient to introduce the following quantities 
 \beqn\label{Yfull}
 \bm Y&=& \bm Y_{\text{def}}+\bm Y_{\text{dynamic}} 
 =\Big( \Imag \sfF+ \frac{i}{4\kappa} e^{i\phi},  \Real \sfF+ \frac{1}{4\kappa} e^{i\phi},   \frac{d+i \gamma}{\sqrt2 }\Big)~,
\\
 \mathcal W_{\text{auxiliary}}&=& - \bm \Theta \cdot \bm Y  =  {\mathcal W}_{\text{def}}+ {\mathcal W}_{\text{dynamic}}~,
 \eeqn
where $\bm Y_{\text{dynamic}}$ refers to the auxiliary fields vacuum expectation values (VEV) in \eqref{Yvev}. 

The deformed transformations of the second supersymmetry are given by 
\beqn\label{susytsf}
\tilde \delta X&=&\sqrt 2  i \tilde \epsilon^\alpha   \Big( W_\alpha +i \gamma \theta_\alpha \Big)~,  \\
\tilde \delta  W_\alpha&=&\sqrt 2 i \Big(\frac{1}{2\kappa}  e^{i\phi} \tilde \epsilon_\alpha  {+}  \frac14 \tilde \epsilon_\alpha \bar D^2 \bar X - i (\sigma^\mu \bar {\tilde \epsilon})_\alpha \p_\mu X  \Big) ~.
\eeqn
 One can check that the $\mathcal N=2$ SUSY algebra is not affected by these constant deformations. 
In the presence of deformations, the fermion transformation rules~\eqref{fermiontransf} get modified as
 \be\label{fermionsusydef}
 \delta_{susy} \PBK{\chi_\alpha \\ \lambda_\alpha}
 =\PBK{ -\sqrt2  \sfF & i (d+i\gamma)\\ i(d+i\gamma)  &-\sqrt 2( \bar \sfF  +\frac{1}{2\kappa}e^{i\phi})} 
 \PBK{\epsilon_\alpha \\ \tilde \epsilon_\alpha}
 = -\sqrt2  \PBK{ Y_2 +i Y_1 & -i Y_3  \\  -i Y_3 &Y_2 -i Y_1 } 
 \PBK{\epsilon_\alpha \\ \tilde \epsilon_\alpha}~,
 \ee
 with  $\bm Y=(Y_1, Y_2, Y_3)$ given in \eqref{Yfull}. 

We also introduce the following parametrization of $\bm Y$ \cite{ADcM}:
\be
\bm Y \equiv \Big( \frac{i}{2} (A^2-B^2), -\frac12 (A^2+B^2) , - i\Gamma \Big)~,  
\ee   
so that
\be 
 \mathcal W_{\text{auxiliary}}=-\bm \Theta \cdot \bm Y  =  A^2 \taa + B^2 \tilde \theta\tilde \theta +2\Gamma\theta \tilde \theta ~. 
 \ee
If  $\Gamma=\pm AB$ (or equivalently $\bm Y \cdot \bm Y=0$),  $\mathcal W_{\text{auxiliary}}$ can be diagonalized and becomes a complete square 
\be 
 \mathcal W_{\text{auxiliary}}=( A \theta   \pm B  \tilde \theta)^2~.
  \ee
This means that there is a combination of two supersymmetries    which is left intact and unbroken. It is related to  the partial supersymmetry breaking we are switching to. 

Supersymmetry is preserved (at least partially), if there {exists} a  linear combination of the  fermions  which is invariant  under the supersymmetry transformation:
\be\label{c12}
 \delta_{susy} ( c_1 \chi_\alpha +c_2 \lambda_\alpha)=0~.
\ee
 This is possible if the  transformation matrix is not invertible, namely
 \be
 \det \PBK{ -\sqrt2 \sfF & i (d+i\gamma)\\ i(d+i\gamma)  &-\sqrt 2(  \bar \sfF  +\frac{1}{2\kappa}e^{i\phi})} = 2\sfF( \bar \sfF  {+} \frac{1}{2\kappa}e^{i\phi} ) +(d+i\gamma)^2=0~.
 \ee
 It is easy to see that this is also equivalent to  $\bm Y\cdot \bm Y=0$   with $\bm Y $ given by \eqref{Yfull}.
 In this case, we always  have a residual $\mathcal N=1$ supersymmetry, therefore realizing partial supersymmetry breaking $\mathcal N=2\rightarrow \mathcal N=1$. 
 
The residual supersymmetry can be found as follows.  The coefficients in \eqref{c12} can be solved yielding: 
 \be\label{rc1c2}
 r\equiv \frac{c_2}{c_1}=\frac{i Y_3}{Y_2 -i Y_1}=\frac {Y_2 +i Y_1}{i Y_3}~.
 \ee
Then the unbroken supercharge is the linear combination:
\be
S= c_1 Q+c_2 \tilde Q~.
\ee 
Indeed from the supersymmetry algebra of $Q, \tilde Q$
  \be
  \{ Q_\alpha, \bar Q_{\dot\alpha }\}=2i \sigma^m_{\alpha \dot {\alpha}}\p_m, \qquad
    \{ \tilde Q_\alpha, \bar {\tilde Q}_{\dot\alpha }\}=2i \sigma^m_{\alpha \dot {\alpha}}\p_m, 
    \qquad
      \{ Q_\alpha, \tilde Q_\alpha \} =     \{ \bar Q_\alpha, \tilde Q_\alpha \}=0
  \ee
one can  easily find that $S$ satisfies the $\mathcal N=1$ algebra
    \be
  \{ S_\alpha, \bar S_{\dot\alpha }\}=2 i \sigma^m_{\alpha \dot {\alpha}}\p_m~,
    \ee
provided that $ |c_1|^2+|c_2|^2=1 $. This condition can always be realized by a trivial rescaling of $c_1, c_2$. 
One can  also explicitly verify that 
\be
\delta^S_\epsilon \lambda=\epsilon S\lambda =  \epsilon (c_1 Q+c_2 \tilde Q)\lambda
= \Big( c_1 (Y_2+i Y_1) +c_2(-iY_3) \Big) \epsilon=0~,
\ee
and similarly $\delta^S_\epsilon \chi=0$. 

To conclude,   $\bm Y\cdot \bm Y=0$ provides the criteria for  a  residual $\mathcal N=1$ supersymmetry.

\section{$\mathcal N=2$ duality}

In this section, we will show the electromagnetic (EM)   duality fully at $\mathcal N=2$ level
\footnote{ We would like to thank  E.  Ivanov for drawing our attention to  ref.~\cite{Ivanov:1997mt} where some points in this section were made   using  different language.  }.  The strategy is  {to make} full use of  various ``long"/``short", chiral/antichiral superfields \cite{ADcM}. 
With this formalism, we can  explicitly see that  our deformations are dual to the triplet of  FI parameters for ($\Real \sfF, \Imag \sfF, \frac{\sfD}{\sqrt2}$). So the deformations can be regarded as the magnetic FI terms. 

\subsection{``Long" and ``short" multiplets} 

We begin with the following   $\mathcal N=2$ ``long" chiral-chiral  superfield \cite{ADcM}:
\be
\hat{\mathcal Z}=Y+\sqrt{2} \tilde \theta \chi - \tilde\theta\tilde\theta \Big( \frac14 \bar D^2 \bar Y +\frac{i}{2} \Phi\Big)~, 
\ee
where $Y, \chi_\alpha, \Phi$ are  $\mathcal N=1$ chiral supefields. 
We can then define   the  $\mathcal N=2$ ``short"  antichiral-chiral  superfield:
\be
\mathcal Z=-\frac{i }{2} \Big(   \tilde D^2 \hat{\mathcal Z} -\bar D^2  \bar{\hat{\mathcal Z}} \Big)~.
\ee
In components, it  {reads} 
\be
\mathcal Z=\Phi -\sqrt2 i \bar{\tilde\theta} \bar D L -  \frac14 \bar{\tilde \theta}^2 \bar D^2\bar \Phi~,
\ee
where 
\be
L=D\chi +\bar D \bar\chi
\ee
is a real linear superfield. 

Similarly, we could begin with the  $\mathcal N=2$ ``long"    chiral-antichiral  superfield:

\be
\hat{\mathcal W}=X+\sqrt{2}  \bar{\tilde \theta}\bar \Omega- \bar{\tilde\theta}^2 \Big( \frac14 \bar D^2 \bar U +\frac{i}{2} X  \Big)~, 
\ee
where $U, \bar \Omega_{\dot \alpha}, X$ are chiral: they are annihilated by $\bar D_{\dot\beta}$.  {In particular, $\bar \Omega$ can be written as} 
$ \bar \Omega_{\dot \alpha}=\bar D_{\dot\alpha} \mathbb L$ with $\mathbb L$ a \emph{complex} linear superfield satisfying $\bar D^2\mathbb L=0$. 
One  can then define   the  $\mathcal N=2$ ``short"   chiral-chiral  superfield:
\be
\mathcal W=-\frac{i }{2} \Big( \bar{  \tilde D}^2 \hat{\mathcal W} -\bar D^2  \bar{\hat{\mathcal W}} \Big)~.
\ee
In components, it reads
\be
\mathcal W=X+\sqrt2 i \tilde \theta W -\frac14 \tilde \theta^2 \bar D^2 \bar X~,
\ee
where 
\be
W_\alpha =\bar D_{\dot\alpha} \Big(\frac12 \bar D^{\dot \alpha}\Omega_\alpha -D_{\alpha} \bar \Omega^{\dot \alpha} \Big)
=  \frac12 \bar D^2 D_\alpha(\mathbb L +\bar{ \mathbb L })~.
\ee
This especially implies that $W$ satisfies the standard  {supersymmetric} Bianchi identity $DW=\bar D \bar W$, which in turn enables  us to define the potential associated to $W$, a real superfield  $V$ such that $W_\alpha=-\frac14 \bar D^2 D_\alpha V$ with $V=-2(\mathbb L +\bar{ \mathbb L} )$.

Since both $\hat{\mathcal W} $ and $ {\mathcal Z}$ are chiral-antichiral,  {we  can consider} the following supersymmetric invariant action 
\be
\int d^2 \theta d^2  \bar {\tilde \theta} \;   {\mathcal Z}\hat{\mathcal W}~.
\ee
Similarly we can also construct the following action from two chiral-chiral superfields $\hat{\mathcal Z},  {\mathcal W}$:
\be
\int d^2 \theta d^2 {\tilde \theta} \;   {\mathcal W}\hat{\mathcal Z}~.
\ee
 {Actually, one can show that the two actions above with imaginary couplings are equal}:
\beqn\label{WZequal}
i\int d^2 \theta d^2 \bar{\tilde \theta} \hat{\mathcal W} \mathcal Z+c.c.
 &=&
\frac12   \int d^2 \theta d^2 \bar{\tilde \theta} \hat{\mathcal W}\Big(\tilde D^2 \hat{\mathcal Z} -\bar D^2  \bar{\hat{\mathcal Z}}
 \Big) +c.c. \nonumber
 \\&=&
  \frac12(-\frac14)   \int d^2 \theta d^2  {\tilde \theta} d^2 \bar{\tilde \theta}    \hat{\mathcal W} \hat{\mathcal Z}   
 -  \frac12(-\frac14)   \int d^2 \theta d^2 \bar{  \theta} d^2 {\tilde \theta}   \bar{  \hat{\mathcal W} }\hat{\mathcal Z} +c.c. \nonumber
\\&=&
\frac12   \int d^2 \theta d^2  {\tilde \theta} \Big(\bar{  \tilde D}^2 \hat{\mathcal W} - \bar D^2  \bar{\hat{\mathcal W}} \Big) \hat{\mathcal Z}  +c.c.   \nonumber
\\&=&
i\int d^2 \theta d^2  {\tilde \theta} \hat{\mathcal Z} \mathcal W+c.c.~.
\eeqn
 
\subsection{Without deformation}
 
To establish the {EM} duality,   we  consider the following action:
\be\label{n2action}
S= \int d^2 \theta d^2  {\tilde \theta}\mathcal F(\hat{\mathcal Z})    
 +i   \int d^2 \theta d^2  \bar {\tilde \theta} \;   {\mathcal Z}\hat{\mathcal W} +c.c.~,
\ee
where the prepotential $\mathcal F$ is a holomorphic function.  The duality in $\mathcal N=2$ theories can be    shown by eliminating different set of variables.   
 
\subsubsection {Electric side}
We {first consider the electric side of the theory by integrating} out $\hat{\mathcal W}$. The equation of motion of $\hat{\mathcal W} $ leads to 
\be
\mathcal Z=0,\qquad \Rightarrow \qquad \Phi=0, \quad L=const   ~.
\ee
Actually one can further show that $L=0$ due to the Bianchi identity  $DW=\bar D\bar W$ in   $\hat{\mathcal W}$
\footnote{ \label{Lis0}
This can be shown as follows:
\beqn
  i\int d^2 \theta d^2 \bar{\tilde \theta} \hat{\mathcal W} \mathcal Z+c.c.
 =
i\int d^2 \theta d^2  {\tilde \theta} \hat{\mathcal Z} \mathcal W+c.c.
   \supset  \int d^2 \theta \chi W +c.c.
&=&\int d^2 \theta \chi ^\alpha (-\frac14 \bar D^2 D_\alpha V) +c.c. \nonumber
  \\&=& - \int d^2\theta d^2 \bar \theta V(D\chi +\bar D\bar \chi)
\eeqn
 The equation  of motion of $V$ gives rise to $L=0$. 
}.
Then we redefine  the field $\chi =i Z$ {such that}
\be
DZ-\bar D\bar Z=-iL=0~.
\ee
The chirality and the above standard supersymmetric Bianchi identity dictates that $Z$ is the field strength superfield of a standard vector multiplet.  $\mathcal {\hat Z}$ becomes then the standard (short) $\mathcal N=2$ chiral-chiral superfield describing a vector multiplet.

The original action after integrating out $\hat{\mathcal W}$,  which will be called electric one,   now becomes 
\beqn
S_e &=& \int d^2 \theta d^2  {\tilde \theta}\mathcal F(\hat{\mathcal Z})    +c.c.
 = \int d^2 \theta \Big( \mathcal F'(-\frac14 \bar D^2 \bar Y  )-  \frac12 \mathcal F'' \chi^2\Big)  +c.c. \nonumber
\\ &=&
\int d^2 \theta d^2 \bar \theta  \; \bar Y \mathcal F_Y
 +\frac12 \int d^2 \theta \; \mathcal F'' Z^2 +c.c.~,
\eeqn
where $\mathcal F_Y\equiv \mathcal F'(Y)$. It is then the standard $\mathcal N=2$ action of a vector multiplet with prepotential $\mathcal F$. 

\subsubsection {Magnetic side}

The action \eqref{n2action} can also be written as 
 
\be\label{n2mag}
 S=\int d^2 \theta d^2  {\tilde \theta}\mathcal F(\hat{\mathcal Z})    
 +i   \int d^2 \theta d^2   {\tilde \theta} \;   {\mathcal  W}\hat{\mathcal Z} +c.c.~,
\ee
thanks to the relation \eqref{WZequal}. 
Now, we would like to integrate out  $\hat{\mathcal Z}$, whose equation of motion yields 
\be\label{WFp}
  \mathcal W=i \mathcal F'(\hat{\mathcal Z}) ~.
\ee
Then the action  takes the form 
\be
S=\int d^2 \theta d^2 \tilde\theta \Big( \mathcal F(\hat{\mathcal Z})  -  \hat{\mathcal Z} \mathcal F'  \Big) ~.
\ee
The integrand is nothing but the Legendre transformation of $\mathcal F$. 

 From  \eqref{WFp}, one could find its inverse function 
\be
\hat{\mathcal Z}=-i\mathcal H'(  \mathcal W) ~,
\ee
such that
\be
 \mathcal F(\hat{\mathcal Z})  -  \hat{\mathcal Z} \mathcal F'  =\mathcal H(\mathcal W)~.
\ee
The construction is reminiscent of  the relation of Lagrangian and Hamiltonian formulation  in classical mechanics once we make the analogy: $-i \mathcal W \leftrightarrow p, \hat{\mathcal Z} \leftrightarrow  \dot x,  \mathcal F\leftrightarrow   L, -\mathcal H \leftrightarrow  H$.
So the dual magnetic theory now becomes 
\be
S_m= \int d^2 \theta d^2  {\tilde \theta}\mathcal H( {\mathcal W})    +c.c.~.
\ee

For clarity, we would also like to write the magnetic theory in terms of components. We expand the action \eqref{n2mag} in terms of $\mathcal N=1$ superfields: 
\be
 \int d^2 \theta d^2  {\tilde \theta}\mathcal F(\hat{\mathcal Z})    
 +i   \int d^2 \theta d^2  {\tilde \theta} \;   {\mathcal W}\hat{\mathcal Z} +c.c.
 = \int d^2 \theta \Big(
  \mathcal (\mathcal F'+i X)  (-\frac14 \bar D^2 \bar Y -\frac{i}{2} \Phi ) 
  -\frac{i}{4} Y \bar D^2 \bar X
  - \frac12 \mathcal F''\chi^2+\chi W 
 \Big) ~.
\ee
We integrate out $\Phi, \chi$:
\be
\delta \Phi: \qquad  X= i \mathcal F'(Y)~,
\ee
\be
\delta \chi:\qquad \chi_\alpha=\frac{W_\alpha}{\mathcal F'' (Y)}~.
\ee
Substituting them back into the action, we obtain}  the magnetic action
\beqn
S_m&=&\int d^2 \theta d^2  {\tilde \theta}\mathcal F(\hat{\mathcal Z})    
 +i   \int d^2 \theta d^2  \bar {\tilde \theta} \;   {\mathcal W}\hat{\mathcal Z} +c.c.
=\int d^2 \theta \Big(
  -\frac{i}{4} Y \bar D^2 \bar X
-  \frac12 \mathcal F''\chi^2+\chi W
 \Big) \nonumber
 \\&=&  \int d^2 \theta  d^2\bar \theta  \;  \bar Y   { \mathcal F}_{ Y}+  \frac12 \int d^2 \theta   \frac{W^2}{ \mathcal F''}+c.c.~.
\eeqn
 We now  define  a new function $\mathcal H$  such that
 \be
X=i \mathcal F'(Y)  ~, \qquad \mathcal H'(X) = i Y~.
\ee
Then it is obvious  to see
\be
 \mathcal F'' \mathcal H'' =\frac{d  \mathcal F'}{dY} \frac{d \mathcal H'}{dX}
 =\frac{ i dX}{dY}\frac{-i dY}{dX}= 1~.
\ee
This enables us to rewrite the magnetic action as
\be
S_m= \int d^2 \theta d^2  {\tilde \theta}\mathcal H( {\mathcal W})    +c.c.
=
 \int d^2 \theta  d^2\bar \theta  \;  \bar X   { \mathcal H}_{ X} +  \frac12 \int d^2 \theta   \mathcal H'' W^2+c.c.~.
\ee
 
The form $S_m$ matches exactly with the form of the original electric theory $S_e$. Thus, the electric theory with chiral scalar $Y$ and prepotential derivative  $ i\mathcal F_Y(Y)$ is dual/equivalent to the magnetic theory with chiral scalar $Y^D=X=i \mathcal F_Y(Y)$ and prepotential derivative $ i \mathcal F^D_{Y^D} (Y^D) =i \mathcal H_X(X)= -Y$.
This establishes the EM duality  at fully   $\mathcal N=2$ level.

\subsection{With   deformation}

 We now turn to adding the deformations and consider the  modified actions as follows
\be\label{defaction}
S= \int d^2 \theta d^2  {\tilde \theta}\mathcal F(\hat{\mathcal Z}-\sqrt2 \theta \tilde \theta \gamma)    
 +i   \int d^2 \theta d^2  \bar {\tilde \theta} \;   ({\mathcal Z}+ \frac{i}{\kappa}e^{i\phi} )\hat{\mathcal W} +c.c.~.
\ee
The dual of the deformations can be found in a similar fashion as above. 

\subsubsection*{$\bullet$ Electric theory}
We can first  integrate out $\hat{\mathcal W}$:
\be
\mathcal Z+ \frac{i}{\kappa}e^{i\phi} =0,\qquad  \Rightarrow\qquad  \Phi=-\frac{i}{\kappa} e^{i\phi}, \quad L=const  \in  \mathbb R~.
\ee
Using the same argument as in footnote \ref{Lis0}, one further finds that $L=0$. 
 {Defining} $\chi_\alpha= i (Z_\alpha -i \theta_\alpha\gamma)$, we have
 \be
\hat{\mathcal Z} -\sqrt2 \theta \tilde \theta \gamma  
=Y+\sqrt{2} i \tilde \theta Z- \tilde\theta\tilde\theta \Big( \frac14 \bar D^2 \bar Y +\frac{i}{2} \Phi\Big)~, 
\ee
where $Z$ satisfies the constraint: 
\be\label{BianchiZ}
DZ-\bar D\bar Z=-4i \gamma  ~.
\ee
This is the modified Bianchi identity of $Z$.    Note that $\mathcal Z$ is not affected by $\gamma$. 

 {One can now obtain the electric action as}
\beqn
 S_e \!=\! \int\! d^2 \theta d^2  {\tilde \theta}\mathcal F(\hat{\mathcal Z}-\sqrt2 \theta \tilde \theta \gamma)    +c.c.
 & \!\!\!\!=\!\!\!\! & \!\int\! d^2 \theta \Big( \mathcal F'(-\frac14 \bar D^2 \bar Y -\frac{i}{2} \Phi ) + \frac12 \mathcal F'' Z ^2\Big) +c.c.\nonumber\\
& \!\!\!\!=\!\!\!\! &
\!\int\! d^2 \theta d^2 \bar \theta \; \bar Y \mathcal F_Y  
+ \frac12  \!\int\! d^2 \theta\;  \mathcal F'' Z^2
 -  \frac{1}{2\kappa} e^{i\phi} \!\!\int\! d^2 \theta \mathcal F_Y  +c.c.~,  \qquad\quad   \label{defSe} 
\eeqn
where $Z$ satisfies the generalized Bianchi identity \eqref{BianchiZ}. 



\subsubsection*{$\bullet$ Magnetic theory}

Using the identity \eqref{WZequal}, the deformed  action  \eqref{defaction} can be written as 
\beqn
S&\!\!\!\!=\!\!\!\!& \int d^2 \theta d^2  {\tilde \theta}\mathcal F(\hat{\mathcal Z}-\sqrt2 \theta \tilde \theta \gamma)   
+i   \int d^2 \theta d^2   {\tilde \theta} \;   {\mathcal  W}\hat{\mathcal Z}  
 - \frac{1}{\kappa}e^{i\phi}  \int d^2 \theta d^2  \bar {\tilde \theta} \;    \hat{\mathcal W} +c.c. \nonumber
\\&\!\!\!\!=\!\!\!\!&
 \int d^2 \theta d^2  {\tilde \theta}\mathcal F(\hat{\mathcal Z}' )   
+i\!   \int d^2 \theta d^2   {\tilde \theta} \;   {\mathcal  W}\hat{\mathcal Z} '
+i\!   \int d^2 \theta d^2   {\tilde \theta} \; \sqrt2 \theta \tilde \theta \gamma   {\mathcal  W} 
 - \frac{1}{\kappa}e^{i\phi}\!\!  \int\! d^2 \theta d^2  \bar {\tilde \theta} \;    \hat{\mathcal W} +c.c.~, \quad
\eeqn
where we have trivially {shifted} the argument of $\mathcal F$. 
The first two terms can be treated as before and we arrive at the magnetic theory:
\beqn 
S_m&=&  
 \int d^2 \theta  d^2\bar \theta  \;  \bar Y   { \mathcal F}_{ Y}+  \frac12 \int d^2 \theta   \frac{W^2}{ \mathcal F''}
 +\gamma \int d^2 \theta d^2 \bar \theta \; \theta^2\tilde \theta^2 \sfD
 + \frac{ i}{2\kappa} e^{i\phi}   \int d^2  \theta X
 +c.c. \nonumber
\\&=&
 \int d^2 \theta  d^2\bar \theta  \;  \bar X   { \mathcal H}_{ X} +  \frac12 \int d^2 \theta   \mathcal H'' W^2
 + 2\gamma \int d^2 \theta d^2 \bar \theta \; V
  + \frac{ i}{2\kappa} e^{i\phi}   \int d^2  \theta X
 +c.c. ~.  \label{defSm}
\eeqn 
 {Therefore} the magnetic theory now {contains} a triplet of FI terms:  
\be
  2\gamma \int d^2 \theta d^2 \bar \theta \; V
  + \frac{ i}{2\kappa} e^{i\phi}   \int d^2  \theta X
 +c.c.
 =2\gamma \sfD +\frac{1}{\kappa}  \sin \phi   \Real \sfF+\frac{1}{\kappa}   \cos \phi \Imag \sfF
 =-4i \bm Y \cdot \bm Y_{\text{def}}~.
\ee

Comparing the  two actions  \eqref{defSe} and \eqref{defSm},
we   clearly see the  duality between deformations and triplet of  {FI couplings}: { $X\leftrightarrow \mathcal F_Y$ and  modification of Bianchi identity  $DZ-\bar D\bar Z=-4i \gamma     $   $\leftrightarrow$  FI D-term $\gamma \sfD$ 
 \footnote{More details of this deformed vector multiplet will be discussed elsewhere \cite{AJD}.}. 

\section{Generalized APT model}

In this section, we discuss the  Antoniadis-Partouche-Taylor (APT) model and its generalizations  with all  deformations we introduced above.  
We will analyse the general $\mathcal N=2$ action based on an arbitrary deformed vector superfield. By computing   the scalar potential, we find that the only non-trivial minima break supersymmetry partially  from $\mathcal N=2\to \mathcal N=1$. 

\subsection {APT model}

In this subsection, we will review the   APT model \cite{APT} which describes  the partial supersymmetry breaking $\mathcal N=2\rightarrow \mathcal N= 1$. 

The  starting point is  {an}  $\mathcal N=2$ chiral-chiral  {superfield introduced} in section \ref{sec:n2def} :
\beqn
 {\mathcal W_{\text{new}}} &=& \mathcal W-\frac{1}{2\kappa}e^{i\phi} \tilde \theta\tilde \theta -\sqrt2\gamma  \theta \tilde \theta 
=X +\sqrt{2} i \tilde  \theta  W  -\frac14 \tilde \theta \tilde \theta \Big(   \bar D^2 \bar X  + 4m\Big), \quad m\equiv\frac{1}{ 2\kappa}  e^{i\phi}~,
 \\
X&=&x+\sqrt{2} \theta \chi -\theta\theta \sfF~.
\eeqn
In this subsection,  we only consider the deformation $\kappa$ and set all others to zero   $\gamma=\phi=0$.
The action of APT model realizing the partial breaking is given by
 \beqn
\mathcal L&=& - i\Big[ \int d^2\theta  d^2 \tilde \theta\;    \mathcal F(\mathcal W_{\text{new}}) - e \int d^2 \theta X  \Big]
-\sqrt2 \xi \int d^4\theta V +c.c. \label{N2action}
\\&=& - i \Big[ \int d^2\theta   \;    \Big(-\frac14  \mathcal F'(X)  (\bar D^2 \bar X+4m )  +\frac12 \mathcal F'' (X)W^2\Big)
- e \int d^2 \theta X  \Big]-\sqrt2  \xi \int d^4\theta V +c.c. \nonumber
\\&=& -i \Big[ \int d^2\theta d^2\bar  \theta   \; \bar X \mathcal F'(X)    
-   \int d^2\theta  \Big(  eX +m\mathcal F'(X)   
-\frac12 \mathcal F''(X) W^2   \Big)  \Big]-\sqrt2  \xi \int d^4\theta V +c.c.~. \qquad
\eeqn
where the holomorphic function $\mathcal F$ is the prepotential and    $m,e,\xi \in \mathbb R $.
As we discussed in the previous section, $eX$ and $m \mathcal F'$ are dual to each other. We add them simultaneously   into the action which is crucial for partial supersymmetry breaking. 
The action can be further rewritten in a compact form as 
\beqn
\mathcal L&=&\int d^4 \theta \mathscr K(X, \bar X)+\int d^2 \theta \mathscr  W(X)+\int d^2 \bar \theta \bar {\mathscr  W}(\bar X)
\\& &  +\Big( \int d^2 \theta \frac{ \mathcal F''(X)}{2} W^2 +c.c.\Big)+2\sqrt2  \xi \int d^4\theta V~,
\eeqn
where the Kahler potential  and superpotential are 
\be
\mathscr K(X, \bar X)= -i\bar X \mathcal F'(X)+ i X \bar{ \mathcal F}'(\bar X), 
\qquad \mathscr  W(X)=  i(eX+m\mathcal F'(X))~.
\ee

 {We now study the scalar potential in order  to find the vacuum of the theory}. Let us first recall the auxiliary fields of various superfields
\be
\qquad W^2 =\taa \sfD^2+..., \qquad \bar D^2 \bar X=4  \bar \sfF+..., \qquad X=x-\taa \sfF+..., \qquad V=\frac12 \taa\taab  {\sfD}+...
\ee
Focusing on the auxiliary field part, the action takes the form 
\beqn
\mathcal L&=& -i \Big[ \int d^2\theta d^2\bar  \theta   \; \bar X X \mathcal F''(x)    
-   \int d^2\theta  \Big(  eX +m X\mathcal F''(x)   
-\frac12 \mathcal F''(x) W^2   \Big)  \Big]-\sqrt2  \xi \int d^4\theta V +c.c +...
\nonumber \\&=& -i \Big[ \tau \sfF \bar \sfF +  {\sfF}(e+m\tau)+\frac12 \tau \sfD^2\Big]-\frac{\sqrt2}{2} \xi\sfD +c.c. +...
\eeqn
where the dots represent terms which do not contain any auxiliary fields and $  \mathcal F''(x) \equiv \tau(x)=\tau_1+i \tau_2 \in \mathbb C $.

 Then, the scalar potential arising from the auxiliary field  is given by 
 \beqn
V(\tau(x))&=&   i \tau \Big(\frac12 \sfD^2 +\sfF\bar \sfF  \Big) + i (m\tau+e)\sfF+\frac{\sqrt2}{2} \xi \sfD +c.c.\nonumber
\\&=&-2\tau_2 (\frac12 \sfD^2 +\sfF\bar \sfF  ) +  i(m\tau +e) {\sfF} - i(m\bar\tau +e) \bar \sfF+ \sqrt{2}\xi \sfD~.
\eeqn
The auxiliary fields  {are  solved using their equations of motion}
\be
\frac{\p V}{\p F} = \frac{\p V}{\p \bar \sfF}=\frac{\p V}{\p \sfD}=0~, 
\ee
with solutions
\be
\sfF= \frac{-i(m\bar \tau +e)}{2\tau_2}~, \qquad  \bar  \sfF= \frac{i(m  \tau +e)}{2\tau_2}~,\qquad \sfD=\frac{\xi}{\sqrt2\tau_2}~.
\ee

Substituting them back, one gets the scalar potential
\footnote{The scalar potential can   be also obtained directly as follows $V=V_D+V_F$:
\be
V_F= \frac{\p \mathscr  W}{\p X} g^ {X\bar X} \frac{\p \bar {\mathscr  W}}{\p \bar X}  =\frac{|m\tau +e|^2}{ 2 \tau_2}~,
\ee
 \be
V_D=\frac{g^2}{8} \Big(2 \sqrt 2 \xi \Big)^2=\frac{\xi^2}{ 2\tau_2}~,
\ee
where $g^ {X\bar X} =(g_ {X\bar X})^{-1}=(\p_X\p_{\bar X} \mathscr  K)^{-1} $ and the real part of the gauge coupling $\frac{1}{g^2}= \Real (-2i\tau)= 2\tau_2$. 
}
\be
V=\frac{|m\tau+e|^2+\xi^2}{ 2 \tau_2}~.
\ee

To find  the vacuum, namely the minimum of the scalar potential  $V(\tau(x))$, we need to extremize  with respect to the scalar field $x$. Equivalently, assuming $\frac{\p\tau(x)}{\p x}\neq 0$, we can  extremize  with respect to  $\tau_1, \tau_2$ and get the following solutions
\be
\tau_1=-\frac{e}{m}~, \qquad \tau_2=\pm\frac{\xi}{m}~.
\ee
One of them is a  discarded by positivity of the kinetic term. The stable vacuum is given by 
\be
\tau_1=-\frac{e}{m}~, \qquad \tau_2=\Big| \frac{\xi}{m}\Big|~. 
\ee

So  the VEV of the auxiliary fields are
\be
\sfF=\bar \sfF=-\frac{m}{2}, \qquad \sfD=\frac{ m \sgn (m \xi) }{\sqrt2} ~,
\ee
and the vacuum potential energy is 
\be
V= |m\xi|~.
\ee
From  previous discussions  \eqref{fermionsusydef}, we easily  find that the fermions transform in the following way:
\beqn
\tilde \delta \lambda  &=&-\sqrt2 \tilde \epsilon  (\bar \sfF+m )=-\frac{1}{\sqrt2} m \tilde \epsilon, \qquad
\tilde \delta \chi  = i \tilde\epsilon  \sfD =i\frac{ m \sgn (m \xi) }{\sqrt2} \tilde \epsilon  ,  \\
  \delta \lambda  &=& i    \sfD \epsilon =i\frac{ m \sgn (m \xi) }{\sqrt2}   \epsilon  , \qquad
 \delta \chi  = -\sqrt2 F \epsilon    =\frac{ m   }{\sqrt2} \epsilon   ~.
\eeqn
It  is then easy to see that
\be
  \delta_{susy} (\lambda+i \sgn(m\xi) \chi)=0~,
\ee
so that a linear combination of two supersymmetries is preserved, and thus the $\mathcal N=2$  supersymmetry is  only partially broken.

\subsection{Generalization of APT model}

As we emphasized, the crucial point in APT model is the simultaneous turning on of electric coupling $eX$ and magnetic coupling $m \mathcal F_X$. Since  in the previous {sections we} found  three deformation parameters, it is natural to generalize the APT model by adding  electric  and magnetic couplings corresponding to the three deformations. 

 The action is almost the same as before:
 \beqn
\mathcal L&=&-  i\Big[ \int d^2\theta  d^2 \tilde \theta\;    \mathcal F(\mathcal W_{\text{new}}) - e \int d^2 \theta X  \Big]
- \sqrt2 \xi \int d^4\theta V +c.c. \nonumber
\\&=&- i \Big[ \int d^2\theta d^2\bar  \theta   \; \bar X \mathcal F'(X)    
-   \int d^2\theta  \Big(  eX +m\mathcal F'(X)   
-\frac12 \mathcal F''(X) W^2   \Big)  \Big]-\sqrt2  \xi \int d^4\theta V +c.c. \qquad\quad
\eeqn
but  now we allow complex $m=m_R+i m_I, \sfD=d+i\gamma$ with $m_R,m_I, \gamma, \xi,e \in \mathbb R$. Note that $e$ is taken to be real since its phase can be absorbed by a rescaling of $X$.

The scalar potential is given by
\beqn
V &=&   i \tau \Big(\frac12 (d+i\gamma)^2 +\sfF\bar \sfF  \Big) + i (m\tau+e)  {\sfF}  + \frac{\sqrt2}{2} \xi (d+i\gamma)+c.c. \nonumber
\\&=&- 2\tau_2 (\frac12 (d^2-\gamma^2) +\sfF\bar \sfF  ) -  2\tau_1 d \gamma + i(m\tau +e)\sfF - i(\bar m\bar\tau +e) \bar \sfF+ \sqrt2 \xi d~.
\eeqn
The auxiliary fields can be solved:
\be
\sfF= \frac{-i( \bar m\bar \tau +e)}{2\tau_2}, \qquad  \bar  \sfF= \frac{i(m  \tau +e)}{2\tau_2},\qquad d=\frac{\xi-\sqrt2 \gamma\tau_1}{\sqrt2\tau_2}~,
\ee
leading to  {the scalar potential}
\be
V=\frac{|m\tau+e|^2+\xi^2-2\sqrt2 \xi \gamma\tau_1+2\gamma^2(\tau_1^2 +\tau_2^2 )}{ 2 \tau_2}~.
\ee

The vacuum sits at 
\be
\tau_1= \frac{ - em_R  +\sqrt2 \gamma \xi }{|m|^2+2 \gamma^2}, \qquad 
 \tau_2=\frac{\sqrt{ (\sqrt2 e \gamma+m_R \xi)^2+m_I^2(e^2+\xi^2) } } {|m|^2+2 \gamma^2}~,
\ee
with auxiliary field VEVs
 \beqn
\bar \sfF&=&\frac{2i e\gamma^2 +i \sqrt{2} m \gamma \xi +e m m_I-  m \sqrt{ (\sqrt2 e \gamma+m_R \xi)^2+m_I^2(e^2+\xi^2) } }
{2\sqrt{ (\sqrt2 e \gamma+m_R \xi)^2+m_I^2(e^2+\xi^2) } }~,
\\   \sfF&=&\bar \sfF^*~,
 \\  d&=&\frac{2 e m_R \gamma +\sqrt2 |m|^2 \xi }
{2\sqrt{ (\sqrt2 e \gamma+m_R \xi)^2+m_I^2(e^2+\xi^2) } }~.
\eeqn
 {One can verify that for the auxiliary field  VEVs above, the following equality always holds:}
\be
\bm Y\cdot \bm Y=0 ~.
\ee
Based on the arguments elaborated in  subsection \ref{susybreaking}, this implies that  there is always a residual $\mathcal N=1$ supersymmetry. 
   
  \subsection{More $U(1)$s towards the complete breaking of supersymmetry}

As we have {just seen, a theory with only one $U(1)$ always has an} $\mathcal N=1$ supersymmetric vacuum, independent of the  {FI parameters} and deformations\footnote{We exclude the singular points $\tau_2=0$ or infinity of runaway behavior and  the trivial case of a free theory with quadratic prepotential. }.  {Hence} it seems impossible to break  {completely} the supersymmetry. However, note that although $\mathcal N=1$ is always preserved, the residual supersymmetry, as a linear combination of the  two   original supersymmetries in $\mathcal N=2$, depends on the deformations and {FI parameters}.  {Therefore if the theory contains} two or more $U(1)$s  {with different residual supersymmetries}, the full system breaks supersymmetry completely. Of course, the different sectors should communicate through matter (not necessarily charged) or gravitational interactions.

More specifically,  consider the   Lagrangian  with   two decoupled $U(1)$s
 
\be
\mathcal L  =\mathcal L^{(1)}+\mathcal L^{(2)}~.
\ee
The previous analysis  applies individually to these two subsectors.   

\be
\bm Y^{(1)}=\bm Y_{\text{def}}^{(1)} +\bm Y_{\text{vev}} ^{(1)}~,\qquad
\bm Y^{(2)}=\bm Y_{\text{def}}^{(2)} +\bm Y_{\text{vev}} ^{(2)}~.
\ee 
 {The full system is thus} characterized by 
\be
\bm Y=\bm Y^{(1)}+\bm Y^{(2)}~. 
\ee
As we have seen in the last subsection, we always have 
\be
\bm Y^{(1)} \cdot \bm Y^{(1)}=\bm Y^{(2)} \cdot \bm Y^{(2)}=0~. 
\ee
However, as long as the two vectors are not aligned $\bm Y^{(2)} \neq c \bm Y^{(1)}$\footnote{ This is true  generically  in the parameter space.}, we immediately have
\be
\bm Y  \cdot \bm Y \neq 0~,
\ee
 {meaning} that $\mathcal N=2$ supersymmetry is broken completely.


 \section{Deformed Dirac-Born-Infeld action}
 
 In this section, we will impose a nilpotent constraint on the deformed $\mathcal N=2$ vector multiplet, which renders  one supersymmetry non-linearly realized. The resulting action is a generalized supersymmetric Dirac-Born-Infeld (DBI) action. We will first study the bosonic part of the  action and find {that it is}  almost identical to the standard bosonic DBI up to some renormalization of coupling constants. This is quite similar to the case of  DBI+FI model where   {the FI parameter only} renormalizes the coupling of the bosonic DBI \cite{ADtM}.  
 In order to differentiate the deformed DBI from the DBI+FI model, we also study the fermionic part using the non-linear SUSY formalism \cite{CFTnonlinear}. 
%
 
 We then study  {SUSY breaking} in our model and  find  {again that there is} always a residual $\mathcal N=1$ supersymmetry  independently of the deformation parameters. However, this unbroken $\mathcal N=1$ supercharge, as a linear combination of $\mathcal N=2$ supercharges, depends on the deformation parameters.

\subsection{Nilpotent constraint on $\mathcal N=2$}

The supersymmetric DBI action  arises from the partial supersymmetry breaking of  $\mathcal N=2\rightarrow \mathcal N=1$.  It was first constructed through the coset method by Bagger and Galperin \cite{BG}.  In   \cite{Rocek:1997hi}, Rocek and Tseytlin found the same action through a nilpotent constraint on the $\mathcal N=2$ superfield. We will thus follow this   elegant nilpotent construction and discuss the deformed DBI.

\subsubsection{Without phase deformation} 
Following    \cite{Rocek:1997hi}, we break  $\mathcal N=2$ by assuming  {the presence of} a Lorentz invariant condensate  $\EV{\mathcal W}=\mathcal W_{\text{def}}  \neq 0$, so 
\be
\mathcal W\rightarrow  {\mathcal W_{\text{new}}}=\EV{\mathcal W}+\mathcal W= \mathcal W+\mathcal W_{\text{def}}~,
\ee
\be
 {\mathcal W_{\text{new}}}= X +\sqrt{2} i \tilde  \theta  W  -\frac14 \tilde \theta \tilde \theta \Big(   \bar D^2 \bar X  + \frac{2}{ \kappa}\Big)~,
\ee
where the deformation $\gamma$ is implicit in $W$. 
We then impose the nilpotent constraint to obtain the non-linearized supersymmetry
\be\label{N2constraint}
 {\mathcal W_{\text{new}}}^2=0~,
\ee
which implies 
\be\label{XWeq}
\frac1\kappa X=WW-\frac12  X\bar D^2 \bar X~.
\ee 
This constraint can be solved to eliminate $X$ in terms of $W$ \cite{BG}:
 \be\label{XWconstraint}
 X=\kappa W^2 -\kappa^3 \bar D^2 \Big[   \frac{W^2 \bar W^2}{ 1+ \mathcal A +\sqrt{1+2 \mathcal A- \mathcal B^2}}  \Big]~,
 \ee
 where we have introduced
 \be\label{AtBt}
 \mathcal A=\frac{\kappa^2}{2} (D^2 W^2+\bar D^2 \bar W^2 )= \mathcal A^*, \qquad 
 \mathcal B=i\frac{\kappa^2}{2} (D^2 W^2-\bar D^2 \bar W^2 )=  \mathcal B^*~.
 \ee
and denote their lowest components as 
 \be
 A= \mathcal A |_{\theta=0}, \qquad  B= \mathcal B |_{\theta=0}~.
 \ee

Before imposing the constraint \eqref{N2constraint}, the most general $\mathcal N=2$ supersymmetric two-derivative action is given in \eqref{N2action}, depending on a prepotential and implemented by two (electric) FI-terms which are linear in the $\mathcal N=1$ superfields $X$ and $V$. After imposing the nilpotent constraint, the prepotential becomes linear in the $\mathcal N=2$ superfield which gives vanishing contribution upon integration over the chiral superspace, and one is left only with the two FI-terms leading to the DBI action and the standard FI-term. 
The DBI action arises from the term linear in $X$:
\be\label{DBI}
 \mathcal L=\frac{1}{4 \kappa g^2  } \Big(\int d^2 \ta X+\int d^2 \tab \bar X \Big)~.
 \ee


  More generally, we can also consider a complex coupling constant
 \be\label{DBI}
\mathcal L=\frac{1}{8\pi \kappa}\Imag   \Big(\tau \int d^2 \theta  X \Big) ~,
 \ee
 where 
\be
\tau=\frac{4\pi i}{g^2} +\frac{\theta }{2\pi}~.
\ee
In the absence of $\theta$-angle and $\gamma$ deformation,  the above action gives rise to the standard   DBI. 
 
\subsubsection{With phase  deformation} 
In the presence of a phase $\phi$,  {eq.~\eqref{DBI} is  modified to} 
 
\be
\frac1\kappa e^{i\phi} X=WW-\frac12  X\bar D^2 \bar X~.
\ee 
 {Nevertheless we can absorb the phase into $X$ by defining $\tilde X=e^{i\phi} X $:
\be
\frac{1}{\kappa}\tilde X=\frac1\kappa (e^{i\phi} X)=WW-\frac12 ( e^{i\phi} X)\bar D^2 (e^{-i\phi}  \bar X)=WW-\frac12 \tilde X\bar D^2 \bar{\tilde X}~.
\ee }
 {The solution is then} the same as \eqref{XWeq} except  for the replacement of $X$ with $\tilde X$:
 \be 
 \tilde X=    \Bigg(\kappa W^2 -\kappa^3 \bar D^2 \Big[   \frac{W^2 \bar W^2}{ 1+ \mathcal A +\sqrt{1+2 \mathcal A- \mathcal B^2}}  \Big]\Bigg)~.
 \ee
The action is
  \be 
\mathcal L=\frac{1}{8\pi \kappa}\Imag   \Big(\tau \int d^2 \theta  X \Big) 
=  \frac{1}{8\pi \kappa}\Imag   \Big(\tau \int d^2 \theta   e^{-i\phi} \tilde X \Big) 
=\frac{1}{8\pi \kappa}\Imag   \Big( \tilde\tau \int d^2 \theta \tilde X \Big) ~,
 \ee
 where $\tilde \tau =e^{-i\phi}\tau$.  {Therefore} the effect of a phase deformation in the action is to rotate the phase of the complex coupling constant.   In the following we will consider a general complex coupling constant which by default has incorporated the phase $\phi$ already.

 \subsection{Bosonic part}
 
 %
%

In this subsection, we will work out the bosonic part of our deformed DBI action. It turns out that in spite of the general deformations,   the resulting bosonic action still takes the  well-known  form of the bosonic DBI action. 

To evaluate the action, let us recall   the component expression of the deformed vector multiplet
\be
W_\alpha = -i \lambda_\alpha +  \theta_\alpha \sfD  -  i (\sigma^{\mu\nu}\theta)_\alpha F_{\mu\nu}    +\taa (\sigma^\mu   \p_\mu {\bar \lambda}  )_\alpha,\qquad D=d+i\gamma, \qquad d, \gamma\in \mathbb R~,
\ee
which satisfies the deformed Bianchi identity \eqref{defW}.
Then we can  calculate
\be
W^2=C+\psi\ta +\taa E~,
\ee
with
\beqn
C=-\lambda^2~, \qquad \psi_{\beta}=-2i \sfD \lambda_\beta+2 F_{\mu\nu} \sigma^{\mu\nu} {}_\beta{}^\alpha \lambda_\alpha~, \qquad
E=\sfD^2-\frac12 (F^2+i F \tilde F)-2i \lambda \sigma ^\mu \p_\mu \bar \lambda~,
\eeqn
 where 
 \be
 F^2\equiv F_{\mu\nu}F^{\mu\nu}, \qquad   F\tilde F\equiv F_{\mu\nu}\tilde F^{\mu\nu}=
  \frac12 \epsilon^{\mu\nu\rho\sigma}F_{\mu\nu}F_{\rho\sigma}~.
 \ee
 
%
%
%

%
%
 
 In the pure bosonic case $\lambda=\bar \lambda=0$, we have 
 \be
 W^2=\taa E=\taa\Big[\sfD^2 -\frac12 (F^2+i F \tilde F)\Big]~, \qquad 
\bar  W^2=\taab \bar E=\taab\Big[\bar \sfD^2 -\frac12 (F^2-i F \tilde F)\Big]~.
 \ee
Since in this case $W^2, \bar W^2$ only have  {non-vanishing} $\taa$ component $E, \bar E\neq 0$, $\mathcal A, \mathcal B$ in \eqref{AtBt} can only contribute  {through} their  lowest components:
  \beqn
 A&=&\mathcal A|_{\ta=0}= -2 \kappa^2 (E+\bar E)=2 \kappa^2  \Big( F^2 - 2 (d^2-\gamma^2) \Big) ~ , \\
 B&=&\mathcal  B|_{\ta=0}= -2 i\kappa^2 (E-\bar E)=-2    \kappa^2  \Big( F \tilde F -4 d\gamma \Big)~.
 \eeqn
With these ingredients, we can now calculate 
 \beqn
 \int d^2 \ta X &=& \int d^2 \ta \Big(\kappa W^2 -\kappa^3 \bar D^2 \Big[   \frac{W^2 \bar W^2}{ 1+ \mathcal A +\sqrt{1+2 \mathcal A- \mathcal B^2}}  \Big] \Big)\nonumber
\\
 &=& \kappa \int d^2 \ta  W^2+4 \kappa^3 \int d^2 \ta d^2 \tab  \frac{W^2 \bar W^2}{ 1+ \mathcal A +\sqrt{1+2 \mathcal A- \mathcal B^2}}  \nonumber
\\
 &=& \kappa E +4 \kappa^3    \frac{E\bar E}{ \Big( 1+    A +\sqrt{1+2   A-  B^2} \Big) }  ~.
 \eeqn
We can decompose it into  real and imaginary parts
 \beqn
2\Real \int d^2 \ta X &=&\kappa (E+\bar E) +8 \kappa^3    \frac{E\bar E}{ \Big( 1+ \mathcal A +\sqrt{1+2 \mathcal A-\mathcal  B^2} \Big) \Big|_{\theta=0}}  
 =\frac{1}{2\kappa} \Big[1- \sqrt{1+2A- B^2}   \Big]~,  \qquad\qquad 
 \\
2\Imag\int d^2 \ta X &=&\kappa (E-\bar E)  =  \frac{i}{2\kappa }B~,
 \eeqn
and then  express  the bosonic action as
\beqn
\mathcal L &\!\!\!=\!\!\!&\frac{1}{8\pi \kappa}\Imag   \Big(\tau \int d^2 \theta  X \Big)\nonumber
\\&\!\!\!=\!\!\!&\frac{1}{2g^2 \kappa}\Real   \int d^2 \theta  X  +\frac{\theta}{16\pi^2 \kappa} \Imag  \int d^2 \theta  X  \nonumber
 \\ &\!\!\!=\!\!\!&
    \frac{1}{8g^2\kappa^2} \Big[1- \sqrt{1+2 \mathcal A- \mathcal B^2}   \Big] 
 +\frac{\theta}{64\pi^2 \kappa^2}   B \nonumber
 \\ &\!\!\!=\!\!\!&
    \frac{1}{8g^2\kappa^2} \Big[1- \sqrt{1+4 \kappa^2  \Big( F^2 - 2 (d^2-\gamma^2) \Big) -4  \kappa^4  \Big( F \tilde F -4 d\gamma \Big )^2 }   \Big]-\frac{\theta}{32\pi^2 }    \Big( F \tilde F -4 d\gamma \Big)~. \qquad
  \eeqn
Note   the term $\theta \gamma d$ which is reminiscent of the standard FI term $\xi d$. This  might provide  an alternative realization of  supersymmetry  breaking via deformation and a non-vanishing $\theta$-angle.

 Solving the constraint:
 \be
 \frac{\p S}{\p d}=0~,
 \ee
we get the auxiliary field
 \beqn 
d &=& 
    \frac{2 \gamma   F\tilde F \kappa ^2}{8
   \gamma ^2 \kappa ^2+1}
   -
   \frac{\gamma  g^2
   \theta   \sqrt{ 1+4 \tilde  \kappa^2 F^2 -4 \tilde\kappa^4( F\tilde F)^2} }{2
   \sqrt{2}  \sqrt{\gamma ^2 \kappa ^2
   \left(g^4 \theta ^2+64 \pi ^4\right)+8 \pi
   ^4}}~,
 \eeqn
 where we introduced  a renormalized coupling
  \be
 \tilde \kappa^2= \frac{\kappa^2}{1+8 \gamma^2 \kappa^2}~.
 \ee

Substituting $d$ back, one gets the final bosonic action
  \beqn
\mathcal L &  =&\frac{1 }{8g^2 \kappa^2}
 -\frac{ \theta F\tilde F     }{32 \pi ^2 \left(8 \gamma ^2 \kappa ^2+1\right)}
  -\frac {1}{8 g^2  \kappa  \tilde \kappa } {\sqrt{ 1+\frac{\theta^2 g^4\gamma^2   \tilde \kappa^2}{8\pi^4}}}
   \sqrt{-  \det\Big(\eta_{\mu\nu}+2 \sqrt{2}\tilde  \kappa F_{\mu\nu}\Big) }  ~. \qquad
 \eeqn
 This action  takes the form of a standard bosonic DBI action, except for the couplings which are renormalized by the deformations.
 
 If we set $\theta=0$,   {the action simply reads}
  \beqn
\mathcal L &=& 
 \frac{1 }{8g^2 \kappa^2}  -\frac{ 1 }{8 g^2  \kappa  \tilde \kappa } 
   \sqrt{-  \det\Big(\eta_{\mu\nu}+2 \sqrt{2}\tilde  \kappa F_{\mu\nu}\Big) }  ~.
 \eeqn
 If  {we furthermore} set $\gamma=\theta=0$, it reduces to the conventional  DBI
 \be
\mathcal L=\frac{1}{8 \kappa^2 g^2} \Big[ 1-\sqrt{-  \det\Big(\eta_{\mu\nu}+2 \sqrt{2} \kappa F_{\mu\nu}\Big) }\Big]
=-\frac{1}{4g^2} F_{\mu\nu}F^{\mu\nu}+...~. 
\ee 
It is worth reminding that  in string theory
\be
\kappa=\frac{\pi \alpha'}{\sqrt2}~.
\ee

\subsection{SUSY breaking}

We now investigate supersymmetry breaking of our deformed DBI action. 

\subsubsection{SUSY breaking in standard DBI+FI}

For comparison, let us  first consider the standard DBI +FI model. We also restrict ourselves to the bosonic part
 \beqn
\mathcal L &=&\frac{1}{4 \kappa g^2  } \Big(\int d^2 \ta X+\int d^2 \tab \bar X \Big)
+\frac{\xi}{\sqrt2  } \int d^2\theta d^2 \bar \theta V \nonumber
 \\ &=&
    \frac{1}{8g^2\kappa^2} \Bigg[1- \sqrt{1+4 \kappa^2  \Big( F^2 - 2 \sfD^2 \Big) -4  \kappa^4  \Big( F \tilde F  \Big )^2 }   \Bigg] 
    +\frac{\xi}{\sqrt8}\sfD~.
  \eeqn
The auxiliary field can be solved 
  \be
  \sfD=-\frac{g^2 \xi \sqrt{1+4 \kappa^2  F^2 -4  \kappa^4  \Big( F \tilde F  \Big )^2 }  }{\sqrt8  \sqrt{1+g^4 \kappa^2 \xi^2}}~,
  \ee
 {whose} vacuum expectation value is {given by}
  \be
  \EV{\textsf D }=  -\frac{ {g^2} \xi }{\sqrt8  \sqrt{1+g^4 \kappa^2 \xi^2}}~.
  \ee
This leads to the following bosonic action
  \be
  S=     \frac{1}{8g^2\kappa^2} 
- \frac{\sqrt{1+g^4 \kappa^2\xi^2 }}{8g^2\kappa^2} \sqrt{1+4 \kappa^2  F^2 -4  \kappa^4  \Big( F \tilde F  \Big )^2 }   ~.
  \ee
  Just like the deformations, the {FI parameter} $\xi$ also renormalizes the couplings.

  The fermion transformation are   
 \beqn
\delta_\epsilon \lambda &=& \epsilon Q\lambda =\sqrt2 i Y_3 \epsilon= i\sfD\epsilon~, \nonumber   \\
\tilde \delta_{\tilde \epsilon } \lambda &=& \tilde \epsilon \tilde Q \lambda
=- \sqrt2 (Y_2-i Y_1) \tilde\epsilon= -\sqrt 2( \bar \sfF  +\frac{1}{2\kappa} )\tilde \epsilon   ~,
 \eeqn   
 where $\bar \sfF$ can be solved from the  {constraint} \eqref{XWconstraint} and expressed in terms of $\EV{\sfD}$. 
The left-over supersymmetry {has to} be a linear combination of  {the $\mathcal N=2$ supersymmetries:}
 \be
 S=c_1 Q+c_2\tilde Q~.
 \ee
The  {ratio $r$ of the coefficients is given by eq. \eqref{rc1c2}}
  \be
 r\equiv \frac{c_2}{c_1}=\frac{i Y_3}{Y_2 -i Y_1} 
=-\frac{i g^2 \kappa \xi }{1+\sqrt{1+g^4 \kappa^2 \xi^2}}~.
  \ee
Then indeed the suparsymmetry transformation $S$  leaves the fermion invariant
 \be
 \delta_\epsilon^S  \lambda= \epsilon S   \lambda=0~.
  \ee
The residual supersymmetry can be more compactly written as
  \be
    S_\alpha =\cos\varphi Q_\alpha- i \sin\varphi \tilde Q_\alpha~,
  \ee
with
  \be
  \tan \varphi= \Big|\frac{c_2}{c_1} \Big| =|r|= \frac{g^2 \kappa\xi}{1+\sqrt{1+g^4 \kappa^2 \xi^2}}~.
  \ee
Therefore, the FI term does not break the supersymmetry in the DBI action. Instead, it rotates the supercharge in the $\mathcal N=2$ space. 
 
\subsubsection{SUSY breaking in  deformed DBI}

Now  we study the supersymmetry breaking in the deformed DBI action. 

From \eqref{XWconstraint}, we can solve the auxiliary field  in $X$ in terms of the auxiliary field in $W$
\be
- {\sfF} e^{-i\phi}=\kappa \sfD^2 +4 \kappa^3 \frac{\sfD^2 \bar \sfD^2}{1+a+\sqrt{1+2a-b^2}}~,
\ee
where
\be
\quad \sfD=d+i\gamma~, \qquad \bar \sfD=d-i \gamma~, \qquad
a=-4\kappa^2 (d^2-\gamma^2)~, \qquad b=8 \kappa^2 d\gamma~.
  \ee
More explicitly, the $\sfF$ and $\bar{\sfF}$ solutions are
\beqn
\sfF&=&- e^{-i\phi} \frac{1+8 i \kappa^2 d \gamma -\sqrt{(1-8 d^2 \kappa^2)(1+8 \kappa^2 \gamma^2)}}{4\kappa}~,
\\
\bar \sfF&=&- e^{ i\phi} \frac{1-8 i \kappa^2 d \gamma -\sqrt{(1-8 d^2 \kappa^2)(1+8 \kappa^2 \gamma^2)}}{4\kappa}~.
\eeqn
This enables us to construct the $\bm Y $ vector
\beqn
\bm Y&=&\Big( \frac{\sfF-\bar \sfF }{2i}-\frac{1}{4i\kappa} e^{ i\phi},  \frac{\sfF+\bar \sfF }{2} +\frac{1}{4\kappa} e^{ i\phi},  \frac{d+i \gamma}{\sqrt2}\Big)\nonumber
\\&=&\Big( 
\frac{-\sqrt{(1-8 d^2 \kappa^2)(1+8 \kappa^2 \gamma^2)} \sin\phi +(i -8 d \kappa^2 \gamma)\cos \phi  }{4\kappa}, 
\nonumber \\&&\quad
\frac{\sqrt{(1-8 d^2 \kappa^2)(1+8 \kappa^2 \gamma^2)} \cos\phi +(i -8 d\kappa^2 \gamma)\sin \phi  }{4\kappa} , 
  \frac{d+i \gamma}{\sqrt2} \Big)~.
\eeqn
 {One can easily check that}
\be
\bm Y\cdot \bm Y=0~,
\ee
implying that  there is always a  residual $\mathcal N=1$ SUSY according to our previous arguments. 
However, the following   $SU(2)_R$ invariant quantity is not zero:
\be
  \bm Y \cdot \bm Y^*=\frac{1}{8\kappa^2}+\gamma^2=\frac{1}{8\kappa\tilde \kappa}~.
\ee
This  defines the partial supersymmetry breaking scale of the theory.

The unbroken supersymmetry can also be worked out as before
  \be
    S_\alpha =\cos\varphi Q_\alpha+  \sin\varphi \tilde Q_\alpha~,
  \ee
with
 \be
  \tan \varphi=  |r| = \Big|\frac{iY_3}{Y_2-i Y_1}\Big| =\frac { \sqrt{1+8 \gamma^2 \kappa^2}  - \sqrt{1-8 d^2 \kappa^2}} { \sqrt{1+8 \gamma^2 \kappa^2} + \sqrt{1-8 d^2 \kappa^2}  }~,
\ee
where $d$ is the VEV
 \beqn 
d &=& 
   -
   \frac{\gamma  g^2
   \theta    }{2
   \sqrt{2}  \sqrt{\gamma ^2 \kappa ^2
   \left(g^4 \theta ^2+64 \pi ^4\right)+8 \pi
   ^4}}~.
 \eeqn
Note that all the possible phase factors have already been absorbed into the definition of supercharges. 

Thus, we see that   we  can only partially break the supersymmetry in $\mathcal N=2$. In order to  break the supersymmetry completely, we need to consider multiple DBIs corresponding to several $U(1)$s, just like  what we discussed in the generalized APT model.  In fact, the situation is similar to D-branes in string theory whose low energy effective action (for a single D-brane) is the supersymmetric DBI, where half of the bulk supersymmetries broken by the D-brane are realized non-linearly on the world-volume. When the bulk has $\mathcal N=2$, for instance in type II superstring compactified on a Calabi-Yau threefold, the world-volume theory has one linear and one non-linear supersymmetry, as in our case of study. A constant magnetic field along the internal directions induces an FI term that one would naively expect to break the linear supersymmetry. However, in the absence of other branes or orientifolds, the magnetic field just rotates the direction of linear supersymmetry or equivalently upon T-duality it rotates the brane. In order to realize complete supersymmetry breaking, one has to consider a system of at least two magnetized branes, or equivalently branes at angles in the T-dual version~\cite{Antoniadis:2004pp,Bianchi:2005yz,Antoniadis:2006eu}.

\subsection{Fermionic part}

As we have seen before, the bosonic part of the deformed DBI action takes the standard form of the bosonic DBI action after eliminating the auxiliary field. The only role of the deformations is to renormalize the coupling constants. This is quite similar to the standard DBI+FI model. So  purely from the bosonic sector viewpoint, it seems that our deformed DBI  is the same as the standard DBI+FI model. In order to find a possible difference, we should also analyze the fermionic part of the action. 


The most straightforward way to consider the fermionic  {contributions is  to directly expand the superfields from the \eqref{DBI} action  \cite{AJL}. This is quite tedious and may not be illuminating. Instead, we will follow the non-linear supersymmetry formalism  {presented} in \cite{CFTnonlinear}. {Using} this formalism,  it was found that in  the standard DBI+FI model, the FI parameter generates an extra term besides renormalizing the coupling constants. It is exactly this extra term  that  is responsible for the gauging of   $R$-symmetry  when coupled to supergravity  \cite{Freedman}.   We will use this non-linear supersymmetry formalism to obtain the fermionic part  of the deformed DBI action. A first analysis indicates that the extra term arising from the FI parameter does not appear and all deformations can be absorbed in the parameters of the standard DBI, exactly as for the bosonic part. This suggests  that if we couple the deformed DBI action to supergravity, it may  not be necessary to gauge the $R$-symmetry.
  
  \subsubsection{The non-linear supersymmetry formalism}
 
 Before discussing the fermionic part, let us  first  review the non-linear supersymmetry formalism elaborated in  \cite{CFTnonlinear}. 
 
Consider a Lagrangian of the type 
 \be\label{FX}
 \mathcal L= F^X+\bar F^{\bar X}~,
 \ee
 which transforms as 
 \be\label{Fxtsf}
 \delta \mathcal L=\delta F^X +\delta \bar F^{\bar X}=
 -2i \p_a(\chi  \sigma^a \bar \epsilon F^X)
  -2i \p_a(\epsilon \sigma^a  \bar\chi\bar F^{\bar X})~.
 \ee
Here $\chi_\alpha$ is the goldstino in the chiral basis, transforming in the following way
 \be
 \delta\chi_\alpha =\epsilon_\alpha -2i\chi\sigma ^m \bar\epsilon \p_m\chi_\alpha~.
 \ee
 This ``chiral" goldstino $\chi_\alpha$  is related to the Volkov-Akulov (VA) goldstino  $\psi_\alpha$ via a field redefinition \cite{Samuel:1982uh}.

 Then up to boundary terms we can rewrite  \eqref{FX} as 
 \be
 \mathcal L=\det A_m^a  (\mathcal B+\bar{ \mathcal B})~,
 \ee
 where 
 \be
 \mathcal B=e^{\delta_\epsilon} F^X\Big| _{\epsilon=-\psi}~,
 \ee
 and 
 \be
 A_m^a=\delta ^a_m -i \p_m \psi \sigma^a \bar \psi+i \psi \sigma^a \p_m \bar \psi~.
 \ee
 Note that $\det A\equiv \det A_m^a$ is  just Volkov-Akulov action density of goldstino.

 \subsubsection{Standard DBI+FI}
 
The standard DBI action can be constructed from the nilpotent  $\mathcal N=2$  superfield   $\mathcal W$ 
  \be
 \mathcal L=\frac{1}{4\kappa  g^2}   \Big( \int d^2 \theta X +\int d^2 \bar\theta\bar X \Big)~,
 \ee
 where $X$ is given by \eqref{XWconstraint} with $W$  the standard field strength superfield of a vector multiplet.
 As shown in \cite{Klein:2002vu,CFTnonlinear},    $F^X=-(\frac{1}{2\kappa}+ \frac14  D^2 X|)$ indeed transforms in the proper way  \eqref{Fxtsf}, thus we can apply the above formalism.  The  {Lagrangian can} be rewritten as 
  \be
 \mathcal L=\frac{1}{4\kappa  g^2}   \Big(\frac{1}{\kappa}+  F^X+\bar F^X    \Big)=
 \frac{1}{4\kappa  g^2} \Big(\frac{1}{\kappa}+   \det A_m^a (\mathcal B+\bar{ \mathcal B}) \Big)~,
 \ee
 where 
 \beqn
 \mathcal B+ \mathcal {\bar B}& =& e^{\delta^*_\epsilon} ( F^X +\bar F^X) \Big| _{\epsilon=-\psi}\nonumber
\\& =& e^{\delta^*_\epsilon} \Big[ ( F^X +\bar F^X)_{\text{bosonic}} \Big] \Big| _{\epsilon=-\psi}\nonumber
\\& =& e^{\delta^*_\epsilon}    \frac{1}{2\kappa} \Big[-2+1 - \sqrt{1+4 \kappa^2  \Big( F^2 - 2 \sfD^2\Big) -4  \kappa^4  \Big( F \tilde F   \Big )^2 }   \Big]  \Big| _{\epsilon=-\psi}\nonumber
\\& =&    \frac{1}{2\kappa} \Big[ -1- \sqrt{1+4 \kappa^2  \Big( \mathcal F^2 - 2  \mathcal  D^2\Big) -4  \kappa^4  \Big(  \mathcal F \tilde { \mathcal F}   \Big )^2 }   \Big]   ~.
 \eeqn
Note that in the second equality, we used the property that the gaugino  $\lambda$ transforms as  {$e^{\delta^*_\epsilon} \lambda | _{\epsilon=-\psi}=0$}. Some rules to implement the operation  $e^{\delta^*_\epsilon} $  can be found in \cite{CFTnonlinear}.  We  also  introduced the following quantities:
 \be
\mathcal D=e^{\delta_\epsilon^*}\sfD|_{\epsilon=-\psi}, \qquad \mathcal F_{ab}=(A^{-1})^m_a (A^{-1})^n_b   (\p_m u_n -\p_n u_m ), 
\qquad u_m =A_m^a  e^{\delta_\epsilon^*}v_a|_{\epsilon=-\psi}~,
\ee
where $v_a$ is the $U(1)$ gauge field.  Here $\mathcal D $ should be regarded as the new auxiliary field although it is composite. 
 
 {Therefore} the standard supersymmetric DBI action written in  non-linear supersymmetry formalism  is~\footnote{Note the constant term proportional to $\det A$, in agreement with ref.~\cite{Bellucci:2015qpa} and an updated version of ref.~\cite{CFTnonlinear}. We have checked eq.~\eqref{DBINL} by a direct computation of the DBI action expanded up to terms of dimension eight~\cite{AJL}. }
 \be\label{DBINL}
 S_{\text{DBI}}=    \frac{1}{8g^2\kappa^2} \int d^4 x\; \Bigg[
2-  \det A \Big[1 + \sqrt{1+4 \kappa^2  \Big( \mathcal F^2 - 2 \mathcal D^2\Big) -4  \kappa^4  \Big( \mathcal F  \mathcal {\tilde F}   \Big )^2 }   \Big] 
 \Bigg]~.
    \ee
    
We can further add the FI term in the DBI action 
\be
\mathcal L_{\text{FI}}=\frac{\xi}{\sqrt2} \int d^4 \theta V~.
\ee
The non-linear supersymmetry formalism can also be applied to rewrite the FI-term but in a more involved way than the DBI action. Indeed, upon decomposing the real superfield $V$ into several constrained superfields and making use of their properties, it was shown in \cite{CFTnonlinear} that the FI term can be rewritten as:  
 \be 
\mathcal L_{\text{FI}}=\frac{1}{2  \sqrt2} \xi \det A\cdot \mathcal D
- \frac{  i }{\sqrt2}\xi \det A \cdot \epsilon^{abcd} [(A^{-1})_a{}^n\p_n\psi]\sigma_b [(A^{-1})_c{}^k\p_k\bar\psi] (A^{-1})_d{}^m u_m~.
 \ee
 
 Eliminating the auxiliary field $\mathcal D$, we get 
 \beqn
 S_{\text{DBI+FI}} &=&    \frac{1}{8g^2\kappa^2} \int d^4 x\; \Bigg[
2-     \det A \Big(1 +  \sqrt{1+ g^4 \kappa^2\xi^2 }   \sqrt{1+4 \kappa^2  \mathcal F^2  -4  \kappa^4  \big( \mathcal F  \mathcal {\tilde F}   \big )^2 }   \Big) \Bigg] \nonumber
\\& &  
- \frac{   i }{\sqrt2}   \xi\int d^4 x\;  \det A \cdot \epsilon^{abcd} [(A^{-1})_a{}^n\p_n\psi]\sigma_b [(A^{-1})_c{}^k\p_k\bar\psi] (A^{-1})_d{}^m u_m
 ~. \qquad   \label{FIterm}
 \eeqn
The second line is responsible for $R$-symmetry gauging when coupled to supergravity \cite{CFTnonlinear}: when lifting to supergravity, $ (A^{-1})_a{}^m\p_m\psi^\alpha \rightarrow \hat D_a \psi^\alpha =e_a{}^m D_m \psi^\alpha -\frac{1}{2M_P} \Psi_a^\alpha$+... 
\footnote{Unfortunately we have a clash of notation here. For clarity we use $\psi$ to denote the goldstino and $\Psi$ to denote the gravitino.}
and thus the second line generates the coupling   $-  \frac{i}{4\sqrt2} \frac{\xi}{M_P^2} \epsilon^{klmn}\Psi_k \sigma_l \bar\Psi_m v_n$, indicating the $R$-symmetry gauging in supergravity that makes the gravitino charged under the $U(1)$ of gauge potential $v_n$.  A direct derivation of the above action by expanding the DBI, as well as the deformed one in the next subsection is under way~\cite{AJL}.

  \subsubsection{Deformed DBI}
  Now we turn to our deformed DBI.     
 We would like to use the non-linear formalism introduced in the previous subsections to rewrite  the deformed action \eqref{DBI}
  \beqn
  \mathcal L &=&\frac{1}{8\pi \kappa}\Imag   \Big(\tau \int d^2 \theta  X \Big)
  = \frac{\tau }{16  \pi \kappa i }  (-\frac14 D^2 X|)+c.c.  ~,
  \eeqn
 in terms of the new variables. Assuming that the non-linear supersymmetry formalism applies also in the presence of the $\gamma$-deformation,\footnote{
 Although, naively, it seems that this is indeed the case, a more careful analysis is needed that goes beyond the scope of the present paper. Explicit calculations are ongoing to check this assumption and clarify the difference between our deformed DBI and the DBI+FI actions~\cite{AJL}.
} 
one can show that  
 \be
 F^X= -\frac{\tau }{16  \pi \kappa i } \Big(\frac{1}{2\kappa}+ \frac14  D^2 X|\Big) 
 \ee
   also transforms in the way like \eqref{Fxtsf}. 
   }
   
 {Hence we can still} use the non-linear supersymmetry formalism to rewrite {the Lagrangian} as 
  \beqn
    \mathcal L  
     &=& \frac{1}{4  \kappa^2 g^2 } +F^X+\bar F^X \nonumber
\\ &=& \frac{1}{4  \kappa^2 g^2 }  +   \det A  (\mathcal B+\bar{ \mathcal B})  ~,
  \eeqn
where
 \beqn
 \mathcal B+ \mathcal {\bar B}& \!\!\!\!\!\!\!\!\!\!\!=\!\!\!\!\!\!\!\!\!\!\!& e^{\delta^*_\epsilon} ( F^X +\bar F^X) \Big| _{\epsilon=-\psi}\nonumber
\\& \!\!\!\!\!\!\!\!\!\!\!=\!\!\!\!\!\!\!\!\!\!\!& e^{\delta^*_\epsilon} \Big[ ( F^X +\bar F^X)_{\text{bosonic}} \Big] \Big| _{\epsilon=-\psi}\nonumber
\\& \!\!\!\!\!\!\!\!\!\!\!=\!\!\!\!\!\!\!\!\!\!\!& e^{\delta^*_\epsilon} \Bigg[  -\frac{1}{4  \kappa^2 g^2 }  +    \frac{1}{8g^2\kappa^2} \Big[1- \sqrt{1+4 \kappa^2  \Big( F^2 - 2 (d^2-\gamma^2) \Big) -4  \kappa^4  \Big( F \tilde F -4 d\gamma \Big )^2 }   \Big] \nonumber
\\&  &\qquad   -   \frac{\theta}{32\pi^2 }    \Big( F \tilde F -4 d\gamma \Big)    \Bigg]  \Bigg| _{\epsilon=-\psi}\nonumber
\\& \!\!\!\!\!\!\!\!\!\!\!=\!\!\!\!\!\!\!\!\!\!\!& \!\!
    \frac{1}{8g^2\kappa^2} \Big[ \!\!-\!1\!-\! \sqrt{1+4 \kappa^2  \Big( \mathcal F^2 - 2 (\bm d^2-\gamma^2) \Big) -4  \kappa^4  \Big(  \mathcal  F \tilde { \mathcal F} -4 {\bm d}\gamma \Big )^2 }   \Big]\!-\!\frac{\theta}{32\pi^2 }    \Big( \mathcal  F \tilde{ \mathcal  F} -4 \bm d\gamma \Big)~.  \qquad\quad
 \eeqn
Here {${\bm d}$ defined by}
 \be
{\bm d}=e^{\delta_\epsilon^*}d|_{\epsilon=-\psi}~,
\ee
is the new composite auxiliary field. Since $\gamma$ is a constant number, {it does not get modified:}
\be
\gamma =e^{\delta_\epsilon^*}\gamma|_{\epsilon=-\psi}~.
\ee
 
Then the complete result takes the form
 \beqn
\mathcal L&=& \det A  (\mathcal B+\bar{ \mathcal B})+\frac{1}{4  \kappa^2 g^2 }  
\nonumber    \\&=& \det A  \Bigg(
  \frac{1}{8g^2\kappa^2} \Big[-1- \sqrt{1+4 \kappa^2  \Big( \mathcal F^2 - 2 (\bm d^2-\gamma^2) \Big) -4  \kappa^4  \Big(  \mathcal  F \tilde { \mathcal F} -4 \bm d\gamma \Big )^2 }   \Big] +  \frac{\theta}{8\pi^2 }    \bm d\gamma  \Bigg)  \qquad
  \nonumber    \\& & 
     +\frac{1}{4  \kappa^2 g^2 }   
-   \frac{\theta}{32\pi^2 }     \det A \cdot \mathcal  F \tilde{ \mathcal  F}   ~.
 \eeqn
 The last term is a total {derivative}: using the definition  $\mathcal F_{ab}=(A^{-1})^m_a (A^{-1})^n_b f_{mn}$, {with} $f_{mn}=\p_m u_n -\p_n u_m $ {the standard field strength of $u_n$}, one finds
 \beqn
 \det A \cdot   \mathcal  F \tilde{ \mathcal  F}  &=& \det A   \cdot  \frac12 \epsilon^{abcd}\mathcal F_{ab} \mathcal F_{ cd}
=  \frac12\det A \cdot   \epsilon^{abcd} (A^{-1})^m_a (A^{-1})^n_b  (A^{-1})^k_c (A^{-1})^l_d   f_{mn} f_{kl}
\nonumber  \\&=&  \frac12\det A    \epsilon^{mnkl} \det(A^{-1})   f_{mn} f_{kl}
\nonumber     \\&=&  \frac12    \epsilon^{mnkl}   f_{mn} f_{kl}
=   f\tilde f~.
 \eeqn
 This is a total derivative and thus can be dropped in the spacetime integral.

Eliminating the auxiliary field and dropping the total derivative term $\mathcal F\tilde {\mathcal  F} $, we  get the deformed DBI action expressed in the non-linear supersymmetry formalism:
  \beqn
S&=&
 \frac{1}{8  \kappa^2 g^2 }    \int d^4 x \Bigg[
 2-
 \det A  \Bigg( 1
  +\frac {\kappa}{   \tilde \kappa } {\sqrt{ 1+\frac{\theta^2 g^4\gamma^2   \tilde \kappa^2}{8\pi^4}}}
   \sqrt{-  \det\Big(\eta_{\mu\nu}+2 \sqrt{2}\tilde  \kappa \mathcal  F_{\mu\nu}\Big) }  \Bigg)  \Bigg]~.
    \eeqn
Especially we see that the second term in \eqref{FIterm} does not appear here, suggesting that there is no need to gauge the $R$-symmetry in order to couple to supergravity. Thus this  {case with deformation seems} different from the DBI+FI model.

\section{Conclusion}

In this paper, we considered the general deformations of $\mathcal N=2$ supersymmetry transformations for a vector multiplet. We have shown that they are dual to the triplet of FI parameters under EM duality. We have then studied the effect of the deformations to the general $\mathcal N=2$ two-derivative action with generic prepotential, as well as to the DBI action realizing one of the supersymmetries non-linearly. We computed the scalar potential and showed that for generic FI terms and deformation parameters, the vacuum is always $\mathcal N=1$ supersymmetric. The complete breaking of supersymmetry requires the presence of at least two $U(1)$'s in analogy with the situation of branes at angles in string theory. 

We also showed that the D-deformation induces an FI term proportional to the theta-angle. However, after the elimination of the auxiliary field, all deformations can be absorbed to a redefinition of the DBI parameters (brane tension and coupling constants) at least within the bosonic sector of the theory. This is also the case of the standard DBI + FI action, implying that the FI parameter and deformation are unobservable within the bosonic sector of the theory. This property is reminiscent of a brane rotation in string theory. An important difference, however, seems to appear in the fermionic sector, where it was observed that the FI term leads to an extra contribution to the action written explicitly in the formalism of non-linear supersymmetry~\cite{CFTnonlinear}. Applying this formalism in our case, where the FI term is generated by the deformation via the theta-angle, we do not find any extra contribution. An explicit computation is currently performed to clarify this point~\cite{AJL}. If such a difference indeed persists, an interesting question is to compare the two theories with the effective action of D-branes in the presence of induced FI terms, for instance via internal magnetic fields~\cite{Antoniadis:2004pp,Bianchi:2005yz,Antoniadis:2006eu}. Note that the extra fermionic contribution appears to be related to the gauging of R-symmetry when coupled to supergravity, suggesting that its absence does not require such a gauging for our case.
Another interesting question is to study the effect of the deformation associated to the change of the Bianchi identity at the $\mathcal N=1$ level and its coupling to supergravity~\cite{AJD}.


  %
%
%
\section *{Acknowledgements}
This work was supported in part by the Swiss National Science Foundation, in part by Labex ``Institut Lagrange de Paris'' and in part by a CNRS PICS grant. We would like to thank Jean-Pierre Derendinger, Fotis Farakos, Gabriele Tartaglino-Mazzucchelli   for enlightening discussions.



\end{document}